\documentclass[11pt]{article}

\usepackage{amsmath,amssymb}
\usepackage[paper=a4paper,text={17cm,24.5cm},centering]{geometry}
\usepackage{graphicx}
\usepackage{ulem}
\usepackage{rotating}

\def\be{\begin{equation}}
\def\ee{\end{equation}}

\def\GO{G^{(0)}_{\Delta x}}
\def\G{G_{\Delta x}}
\def\pn{p_n(\Delta x)}
\def\pnO{p_n^{(0)}(\Delta x)}

\DeclareMathOperator{\erfc}{erfc}

\begin{document}

\title{
Particle-number distribution in large fluctuations at the tip of branching random walks
}

\author{A.H. Mueller${}^{(1)}$, S. Munier${}^{(2)}$\\
  \footnotesize\it  (1) Department of Physics, Columbia University,
  New York, NY 10027, USA\\
  \footnotesize\it  (2) CPHT, CNRS, \'Ecole polytechnique, IP Paris, F-91128 Palaiseau, France
}
\date{August 4, 2020}

\maketitle

\begin{abstract}
  We investigate properties of the particle distribution near
  the tip of one-dimensional branching random walks at large times~$t$,
  focusing on unusual realizations in which
  the rightmost lead particle is very far ahead of its expected
  position -- but still within a distance smaller than
  the diffusion radius $\sim\sqrt{t}$. Our approach consists in a study of the
  generating function  $\G(\lambda)=\sum_n \lambda^n \pn $
  for the probabilities $\pn$ of observing $n$ particles in
  an interval of given size $\Delta x$
  from the lead particle to its left, fixing the position of the latter.
  This generating function can be expressed with the help of functions solving
  the Fisher-Kolmogorov-Petrovsky-Piscounov (FKPP)
  equation with suitable initial conditions.
  In the infinite-time and large-$\Delta x$ limits, we
  find that the mean number of particles in the interval grows exponentially
  with $\Delta x$, and that the generating function
  obeys a nontrivial scaling law, depending on $\Delta x$ and $\lambda$
  through the
  combined variable $[\Delta x-f(\lambda)]^{3}/\Delta x^2$, where
   $f(\lambda)\equiv -\ln(1-\lambda)-\ln[-\ln(1-\lambda)]$.
  From this property, one may conjecture that the
  growth of the typical particle number with the size of the interval
  is slower than exponential,
  but, surprisingly enough, only by a subleading factor at large $\Delta x$.
  The scaling we argue is consistent
  with results from a numerical integration of the FKPP equation.
\end{abstract}

\noindent\hrulefill
\setcounter{tocdepth}{2}
\tableofcontents
\noindent\hrulefill

\section{Introduction}

Branching random walks are stochastic processes which rule the time
evolution of a set of particles through
a random walk combined with a branching process.
Each particle is characterized
by its position in some space
and may evolve in time by splitting into several descendants,
and/or by randomly changing its position,
independently of the other particles
present at the same time in the set.
Branching random walk (BRW) processes turn out to be
relevant in many physical and
biological context. Hence, studying the universal properties of
such processes is of wide interest. In this paper, we will
address BRWs in one space dimension,
starting with one single particle,
and we will focus our discussion essentially
on the branching Brownian motion (BBM).
In this context, in each realization at any given time there
are two lead particles, and
the set of the positions of the latter indexed by the time form
two boundaries in the two-dimensional plane.
The distribution of the positions of these boundaries is given by
solutions to the well-known
Fisher-Kolmogorov-Petrovsky-Piscounov (FKPP)~\cite{f,kpp} equation
(For a review, see Ref.~\cite{VANSAARLOOS200329}).

Our motivation for studying BRWs initially
comes from particle physics~\cite{Munier:2014bba}.
In the context of the scattering of hadrons at very high energies,
the state of the hadrons, which determines
observables such as the interaction cross section, is essentially
a set of many gluons, which turn out to be
generated by a peculiar BRW. This picture emerges
from the theory of the strong interactions,
quantum chromodynamics, in the semi-classical limit relevant
when the energy at which the scattering occurs is very large
compared to any other energy scales, such as the hadron masses or
the momenta transfered to the particles that go to the final state.
An example of direct application of some detailed properties of
BRWs in this physical context was found recently.
We showed indeed that a subset of the events observed in
the data collected at high-energy colliders, called ``diffractive
events'' due to the characteristic angular distribution of the particles
measured in the final state,
can be understood as a consequence of the fluctuations of the position
of a lead particle in a 
BRW~\cite{Mueller:2018zwx,Mueller:2018ned}.
The corresponding statistical physics problem is actually the genealogy of the
particles generated by a BRW that have a position larger than some number,
which, in the context of particle physics, is determined by some intrinsic
properties of the interacting particles~\cite{Anh:2018lcc}.

Fluctuations in a BRW mainly occur in two
places: in the beginning of the evolution, when the system
consists of few particles,
and throughout the evolution in the vicinity of
the two lead particles.
Based on this observation and modeling these two kinds
of fluctuations in a very simple way~\cite{Mueller:2014gpa},
it was possible to arrive at
new quantitative results for general BRWs,
such as analytical expressions for
the moments of random variables defined on the
set of particles present in the system at a given time.
Such a phenomenological model has helped to formulate
a mathematical conjecture, that by now has been proven and
generalized~\cite{2018arXiv180605152M}.

The fluctuations occurring in the beginning of the time
evolution have an
effect on the particle distribution
that persists throughout: They
mainly result in an effective global
shift of the mean positions of the particles in the vicinity of
the boundary of the BRW with respect to
the expectation values of their positions in typical events,
by a distance which tends to
a realization-dependent finite constant at large times.
We call them ``front fluctuations''; They had been formalized in mathematics
in the earlier work of Lalley and Sellke~\cite{lalley1987}.
Instead, late-time fluctuations originating from particles
near the boundaries of the BRW
have a small effect on the position of the latter,
that vanishes as they occur at larger times.
We call them ``tip fluctuations''.
In the phenomenological model, in a given realization,
both fluctuations are needed to
send the lead particle significantly ahead of its expected position.

The properties of the tip of BRWs
turn out to provide insight into the free energy
of spin glasses~\cite{DerridaSpohn88},
and have therefore been an active research topic for
some time. Several recent works have focused on the properties of
lead particles in the BBM.
For example, the large deviations of a lead particle have recently
been discussed~\cite{Derrida_2016}, as well as the correlations between the positions
of the two lead particles~\cite{Ramola_2015}.
The systematic study of the properties of particle distributions near the tip of
BRWs was pioneered in the physics literature in
Ref.~\cite{Brunet_2009,Brunet2011};
See also Ref.~\cite{arguin2012,aikedon2013,Arguin2013}
for parallel mathematical developments. These works were essentially focused on
the limiting distribution of extremal points in the BBM, namely, as seen from
the lead particle.
The object of the present paper is to propose an approach
to the study of the statistical properties of the particles
in large tip fluctuations, that is, near the tip of a BRW
{\it conditioned to have its lead particle at a position very far from its expected
position.}
In the phenonomenological model described above,
the main assumption on tip fluctuations is actually
that they consist of low-density
sets of particles sent far away from the expected position of the
lead particles,
in the region which is void of particles in a typical realization.
The main question we want to address
is whether this crucial assumption\footnote{Note that these tip fluctuations
  also drive the fluctuations in the position of stochastic fronts,
  namely determine the statistics
  of the position of the bulk of systems of particles generated by branching random walks
  supplemented with a selection process (see Ref.~\cite{Brunet:2005bz}), as well
  as the statistics of the genealogical trees (see Ref.~\cite{Brunet:2007iqc}).
  Recently, evidence was given that the genealogies of extreme
  particles in BRWs without selection can be explained by the statistics
  of what we call tip fluctuations occurring
  during the evolution~\cite{0295-5075-115-4-40005}.
  } is indeed verified in an actual BRW.
\begin{figure}[t]
  \includegraphics[width=0.5\textwidth]{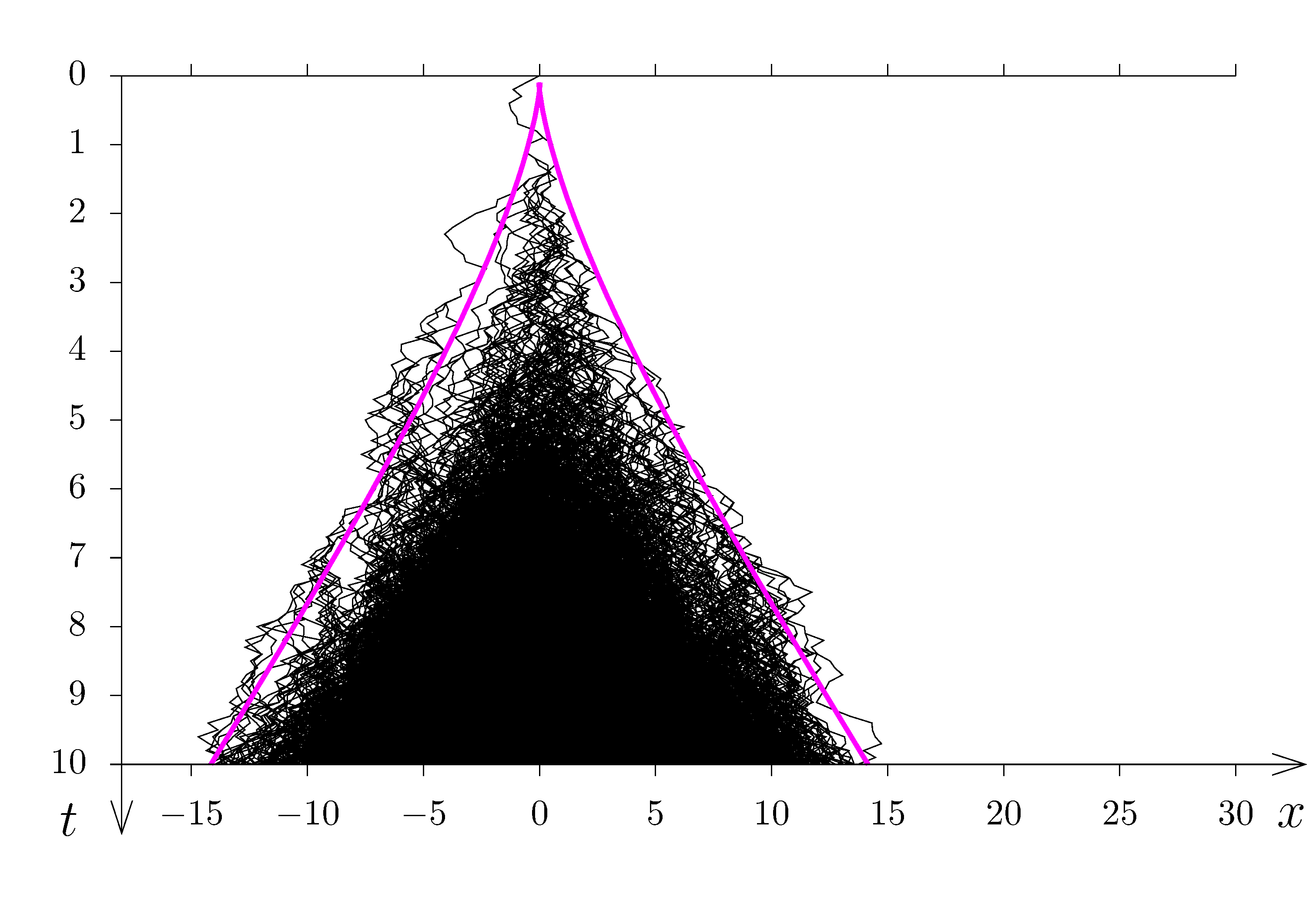}
  \includegraphics[width=0.5\textwidth]{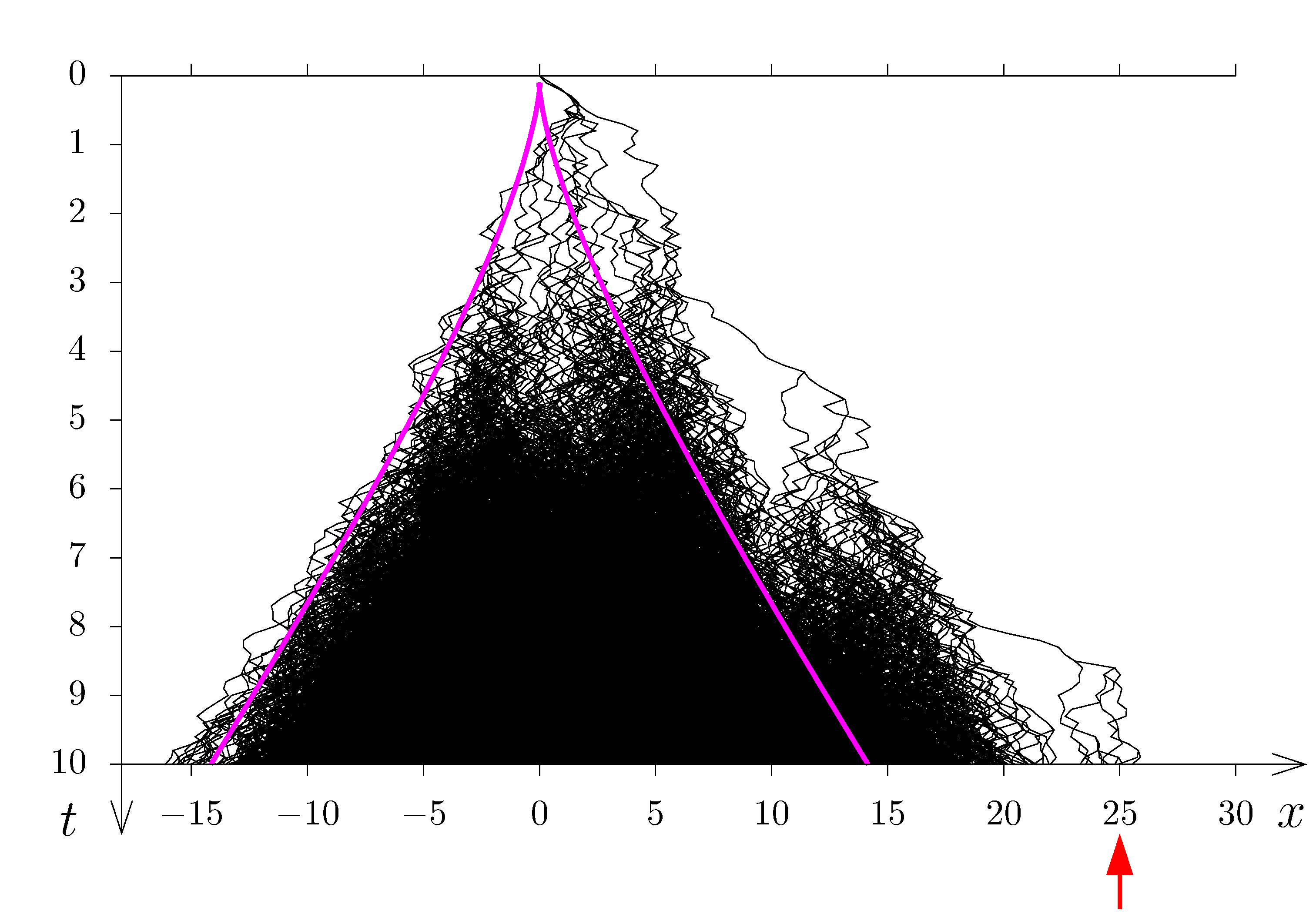}
  \caption{\small Typical (left) and unusual (right) realizations of the branching Brownian
    motion. The continuous lines
    represent the expected values $x=\pm m_t$
    of the positions of the lead particles, extracted from a numerical solution
    to the FKPP equation.
    The realization in the right plot is obtained by conditioning the rightmost particle
    at final time $t=10$ to be to the right of the position $x=25$ indicated
    by the arrow, that is to say, way
    ahead of its expected position. (The probability of such an event
    is about $10^{-4}$).
    We see that while in the typical realization the lead particles 
    are quite close to their expected values throughout, in the
    unusual realization instead, the particles on the
    right of $x=25$ are brought there by a combination of a large front fluctuation, namely
    which develops at relatively early times,
    and a large tip fluctuation, namely which occurs at late times.
    In the latter event, the density of particles in an interval of size say~$\Delta x\sim5$
    from the tip is visibly quite low, much lower than in the same interval
    in a typical realization.
    \label{fig:realizations}
  }
\end{figure}

To this aim, an appropriate quantity to compute is the distribution
$p_n(\Delta x)$ of
the number $n$ of particles in the finite interval $\Delta x$ from
the position of the lead particle. 
(See Fig.~\ref{fig:realizations} for an illustration.)
Generally speaking, quantities like $p_n(\Delta x)$
were addressed in the literature in the case of sets of particles
characterized by a single number chosen in various ways:
For example, the sets of particles drawn from an
inhomogeneous Poisson point process on a line or resulting from a
Ruelle cascade~\cite{Brunet2011,Ruelle87} were considered; or simply
sets of independent and
identically distributed random variables~\cite{Sabhapandit2007};
or, also, sets of eigenvalues of random matrices~\cite{Perret2015}.
In the case in which the generating process is the BBM
(or more generally BRW),
Brunet and Derrida have shown~\cite{Brunet2011}
that all properties of the particle distribution
may be deduced from solutions to the
FKPP equation with peculiar initial conditions, which is obeyed by
generating functions of the particle-number probability distributions.
In this paper, we shall discuss properties of these
generating functions. While at this point exact solutions seem out of reach,
we are able to find an interesting scaling form,
from which one may conjecture qualitative features of the particle distribution
near the tip in typical realizations of rare tip fluctuations.

In Sec.~\ref{sec:general},
we formulate the calculation of the statistics of the number
of particles in the tip of a branching random walk using a generating
function; In Sec.~\ref{sec:shift},
we present our analytical insight on the solutions
of the equations governing the time evolution of the generating function;
In Sec.~\ref{sec:distribution},
we derive a few consequences for the mean of the particle
numbers in the tip, and we propose a conjecture for the
parametric $\Delta x$-dependence of their typical value.


\section{\label{sec:general}General framework}

As announced in the Introduction, we will essentially discuss branching
Brownian motion (BBM) in one dimension. The evolution variable, namely
the time, will be
labeled by $t$, and the line on which the particles
move by~$x$. We shall always start
at $t=0$ with a single particle at the origin $x=0$, which
evolves in time through two independent stochastic processes:
Diffusion (we shall set the diffusion coefficient $D$ to~$1$), and
splitting to two particles (at rate unity), which further evolve
independently according to the very same rules.

We expect the results we will obtain to be universal,
namely they should apply to a larger class of branching random
walk models (such as the ones discussed in Ref.~\cite{Brunet2011}),
up to the replacement of the small set of constants which depend
on the detailed definition of the
particular process we may consider.

Our aim is to understand the particle density in a tip fluctuation.
It is natural to select realizations of the BBM in which the rightmost
particle, at some given large time~$t$, is significantly ahead of its expected
position, and then, to count the particles within an interval of fixed size
$\Delta x$ from the position of that lead particle.
In this section, we shall
introduce the formalism useful for our analytical work, as well
as the numerical tools.

\subsection{Generating functions of particle-number
  probabilities and their evolution}

We start by introducing the joint probability
$Q_n(x,x-\Delta x,t)$ that, at time $t$  (which we will
eventually take infinite), there is no
particle to the right of $x$, and exactly $n$ particles to the right of
$y\equiv x-\Delta x$ ($\Delta x>0$).
Obviously, at the initial time, when the system consists of one single
particle at the origin $x=0$,
\be
Q_n(x,x-\Delta x,t=0)=
\delta_{n,0}
\Theta(x-\Delta x)
+\delta_{n,1}\Theta(x)\Theta(\Delta x-x),
\ee
where $z\mapsto\Theta(z)$ is the Heaviside function.
Following Brunet and Derrida~\cite{Brunet2011}, who built on
Ref.~\cite{doi:10.1002/cpa.3160280302},
we introduce the generating function\footnote{
  We slightly simplify the notations in Ref.~\cite{Brunet2011}:
  Our $\psi_{\lambda}$ would read
  $\psi_{0\lambda}$ in there.}
\be
\psi_{\lambda}(x,x-\Delta x,t)
=\sum_{n=0}^\infty \lambda^n Q_n(x,x-\Delta x,t).
\ee
Then, defining
\be
\phi(x)\equiv \lambda\,\Theta(x)+(1-\lambda)\,\Theta(x-\Delta x),
\label{eq:decomp_phi}
\ee
it is a straightforward calculation to
establish that the two-variable function
\be
H_\phi(x,t)\equiv\psi_{\lambda}(x,x-\Delta x,t)
\label{eq:H=psi}
\ee
obeys the FKPP equation\footnote{On terminology: In the literature,
  by FKPP equation
  is usually meant the nonlinear partial differential equation
  $\partial_tu=\partial^2_xu+u(1-u)$, up to dimensionful positive
  coefficients. We shall also use this name for the equation solved
  by $H_\Phi\equiv 1-u$.
  }
\be
\partial_t H_{\phi}(x,t)=\partial_x^2 H_\phi(x,t)+H_\phi^2(x,t)-H_\phi(x,t)
\ee
with the initial condition
\be
H_\phi(x,t=0)=\phi(x).
\ee
It will prove useful to decompose $\phi(x)$ as the sum
\be
\phi(x)=\phi_A(x)+\phi_B(x)-1,
\ee
where
\be
\begin{cases}
\phi_A(x)\equiv 1+(1-\lambda)[\Theta(x-\Delta x)-\Theta(x)]\\
\phi_B(x)\equiv\Theta(x).
\end{cases}
\ee
In these equations, $\Delta x$ and $\lambda$ are parameters,
the value of which is fixed
as far as the $t$-evolution is concerned.

We then introduce the probability $Q(x,t)$ that there is no particle
to the right of $x$. It is known to also obey the FKPP equation
with the initial condition
\be
Q(x,t=0)=\Theta(x)=\phi_B(x),\quad\text{and therefore,}\quad
Q(x,t)=H_{\phi_B}(x,t).
\label{eq:Q=HphiB}
\ee

Let us now write the generating functions for the
particle-number probabilities within the interval $\Delta x$
from the lead particle, considering two cases:
\begin{itemize}
\item First, the case in which the position
of the lead particle is unconstrained, in order
to make contact with Ref.~\cite{Brunet2011}:
If we call $p_n^{(0)}(\Delta x)$ the probability to find
$n$ particles in the interval of size $\Delta x$ from the lead particle,
then we write the corresponding generating function as
\be
\GO(\lambda)=\sum_{n=1}^\infty\lambda^n p_n^{(0)}(\Delta x).
\ee
(In addition to $\lambda$ and $\Delta x$, $\GO(\lambda)$ also depends on the
time $t$, which is not explicitly mentioned since this dependency goes away
in the limit $t\rightarrow+\infty$ of interest here).

\item Second, for our purpose of understanding the
tip fluctuations in rare realizations in which the lead
particle sits very far ahead of its expected position,
we need to fix its position $x$.
Thus the quantity of interest for us
is the generating function of
the probability $p_n(\Delta x)$ that there be a number~$n$ of particles in the interval
$[y\equiv x-\Delta x,x]$ {\it given} that the lead,
rightmost, particle is at the position~$x$:
\be
\G(\lambda)=\sum_{n=1}^\infty \lambda^n \pn.
\label{eq:G_expansion_pn}
\ee
(Again, $\G(\lambda)$ also depends on the time $t$ and in addition, on $x$,
dependencies which are understood since they are only relevant for finite~$t$,
a case that we shall not investigate here.)
\end{itemize}

The generating functions $\GO(\lambda)$ and $\G(\lambda)$
do not directly obey the FKPP equation,
unlike $\psi_\lambda$, but they can 
easily be deduced from $\psi_\lambda$ and $Q$.
It is clear that $\GO(\lambda)$
is related to $\psi_\lambda$ through the following formula:\footnote{
  Note that the derivative
  ${\partial\psi_\lambda(x',y,t)}/{\partial x'|_{x'=x,\,y=x-\Delta x}}$
  that appears in Eqs.~(\ref{eq:def_G0}) and~(\ref{eq:def_G})
  is with respect to the position of the lead particle, {\it keeping fixed
  the position of the left bound of the interval} in which one counts
  the particles. Therefore, this derivative does not coincide with a
  derivative of the two-variable function $H_\phi(x,t)$ that obeys the FKPP
  equation.
  }
\be
\GO(\lambda)=\int_{-\infty}^{+\infty}dx
\left.\frac{\partial\psi_\lambda(x',y,t)}{\partial x'}
\right|_{x'=x,\,y=x-\Delta x}.
\label{eq:def_G0}
\ee
It is equally straightforward to see that
\be
\G(\lambda)=
\frac{\left.\partial\psi_\lambda(x',y,t)
  /\partial x'\right|_{x'=x,\,y=x-\Delta x}}
     {\partial Q(x,t)/\partial x}.
\label{eq:def_G}
\ee

Note that $\GO(\lambda)$ and
$\G(\lambda)$ can be written as a series of powers of $1-\lambda$
with the help of the factorial moments $n^{(k)}(\Delta x)$ of the particle number $n$
in the interval $[x-\Delta x,x]$. For example,
\be
\G(\lambda)=
\sum_{k=0}^\infty \frac{(-1)^k}{k!}{(1-\lambda)^k}n^{(k)}(\Delta x)
\quad \text{with $n^{(k)}(\Delta x)\equiv \langle n(n-1)\cdots(n-k+1)\rangle$},
\label{eq:G_expansion}
\ee
where the expectation is taken over realizations which have the rightmost lead
particle at the final time at position~$x$.


\subsection{Infinite-time asymptotics of the generating functions}

\subsubsection{Traveling wave solutions to the FKPP equation}

When going to large times,
the solutions to the FKPP equation
$Q$ and $H_\Phi$, with $\Phi\equiv\phi$, $\phi_A$ or $\phi_B$,
all converge to
{\it traveling waves} the shape of which is stable through
time evolution and universal,
i.e. largely independent of the details of the initial
condition~\cite{MR705746}.
We shall call $F$ the function of the space variable
that encodes this shape.
In this large-$t$ limit, the evolution
is essentially a translation of the traveling wave along the $x$-axis. We
shall denote by $m_t$ the position of the wave front of $Q$
at time $t$. The leading $t$-dependence of $dm_t/dt$
is also universal at large $t$.
In practice, $m_t$ can be computed by taking a
particular integral of $Q(x,t)$
over the $x$-variable,
in which case it coincides with the expectation value
of the position of the lead particle. More precisely, if we choose the definition
\be
m_t\equiv\int_{-\infty}^{+\infty}dx\,x\,\frac{\partial Q(x,t)}{\partial x},
\ee
then $m_t=\langle x_{\text{lead particle}}\rangle$, by definition of $Q$.
It is then convenient to introduce a position variable in the frame of the
traveling-wave front:
\be
\xi\equiv x-m_t.
\ee

Equiped with these notations, we can
rewrite the asymptotics of $Q$ as
\be
Q(\xi+m_t,t)\underset{t\rightarrow +\infty}{\simeq} F(\xi).
\label{eq:travelingwaveQ}
\ee
Because of the universality of the shape encoded in the function $F$ and of
the leading-$t$-dependence of $m_t$ at large $t$, 
the solutions for $Q$ and $H_\phi$ can only differ by a shift
$g(\lambda,\Delta x)$ in the argument of $F$. Hence
\be
H_{\phi}(\xi+m_t,t)
\underset{t\rightarrow +\infty}{\simeq} F[\xi-g(\lambda,\Delta x)].
\label{eq:travelingwave}
\ee
The shape function $F(z)$
connects 0 at $z=-\infty$ to 1 at $z=+\infty$. Its asymptotics
read
\be
F(z)\simeq
\begin{cases}
  1-C\times z\, e^{-z}  & \text{for large positive $z$}\\
  C'\times
  e^{(\sqrt{2}-1)z}  & \text{for large negative $z$},
\end{cases}
\label{eq:asymptotics_F}
\ee
where $C$ and $C'$ are constants of order~1, which are unambiguous
(although not calculable analytically) once $m_t$ is properly defined.
The $t$-dependent real number $m_t$
has the following time dependence for large~$t$:
\be
m_t\underset{t\gg 1}{\simeq}2t-\frac32\ln t,
\label{eq:mtBBM}
\ee
up to an additive constant,
independent of $\lambda$ and of $\Delta x$
but dependent on the very definition of the position of the
wave front,
and up to terms which vanish when $t$ is large.
With the definition $m_t\equiv\langle x_{\text{lead particle}}\rangle$,
one has $m_{t=0}=0$; But no analytical expression for $m_t$
is known when
$t$ is small (i.e. of order unity).

Although we will eventually be interested only in
the infinite-time limit, we will need
the shape $F_t(z)$ of the FKPP traveling wave
at finite (but large) time~$t$:
\be
F_t(z)\simeq 1-C \times z\,e^{-z} e^{-z^2/4t}
\quad\text{for large positive $z$}.
\label{eq:FtBBM}
\ee
We note that $F_t(z)\simeq F(z)$ for $z<2\sqrt{t}$,
and that $1-F_t(z)$ goes to 0 very quickly as soon as
$z>2\sqrt{t}$, due to the Gaussian damping factor.


\subsubsection{Generating functions 
  expressed in terms of the shift}

In Ref.~\cite{Brunet_2009},
Brunet and Derrida noticed that some quantitative properties of
the tip of a branching random walk can be deduced from the only
knowledge of the difference in the positions of the traveling
waves resulting from the evolution of different initial conditions
to very large times.
In the context of the present paper, it turns out that
a similar statement can be made:
Generating functions of the particle-number probabilities in an interval
of size $\Delta x$ from the lead particle of a BRW are
related to the differences of the positions of the asymptotic traveling
waves $H_\phi$ and $H_{\phi_B}$, encoded in the shift function $g$
defined in Eq.~(\ref{eq:travelingwave}).

Let us start with Eqs.~(\ref{eq:def_G0}) and~(\ref{eq:def_G}).
The derivative of $\psi_\lambda$ in these formulas
is easily expressed in terms of
derivatives of the shape function~$F$ of the traveling waves
and of the shift function $g$. From the relation between
$\psi_\lambda$ and $H_\phi$, Eq.~(\ref{eq:H=psi}), and from the
asymptotic behavior
of the latter at infinite times,
see Eq.~(\ref{eq:travelingwave}), we deduce that
the generating function~$\GO(\lambda)$ is a simple function
of the shift~$g(\lambda,\Delta x)$:
\be
\GO(\lambda)\simeq 1-\frac{\partial g(\lambda,\Delta x)}{\partial\Delta x}.
\label{eq:G0fromg}
\ee
In the same way, starting this time from Eq.~(\ref{eq:def_G}),
we find the following expression
for the large-time asymptotics of the generating function
$G_{\Delta x}(\lambda)$:
\be
\G(\lambda)\simeq \left[1-\frac{\partial g(\lambda,\Delta x)}
{\partial\Delta x}\right]
\frac{F'[\xi-g(\lambda,\Delta x)]}{F'(\xi)}.
\ee
In order to simplify further this formula, let us pick a value
of $x$ such that  $|\xi|\gg 1$ and $|g(\lambda,\Delta x)|\ll |\xi|$.
Then, using Eq.~(\ref{eq:asymptotics_F}), we see that $\G(\lambda)$
boils down to a function of the shift only:
\be
\G(\lambda)\simeq
\left[1-\frac{\partial g(\lambda,\Delta x)}{\partial\Delta x}\right]
\times
\begin{cases}
e^{g(\lambda,\Delta x)}
& \text{$x>m_t$}\\
e^{-(\sqrt{2}-1)g(\lambda,\Delta x)}
& \text{$x<m_t$}.
\end{cases}
\label{eq:Gfromg}
\ee

It is clear from these formulas that it is enough to compute
$g(\lambda,\Delta x)$ in order to get the infinite-time
asymptotics of $\G(\lambda)$.

\subsection{Particle-number probabilities from the generating function}

The particle-number probabilities $\pn$
(and, similarly, $\pnO$)
can obviously be obtained
from the generating function through an integration in the complex
plane:
\be
\pn=\oint\frac{d\lambda}{2i\pi}\lambda^{-n-1}\G(\lambda),
\label{eq:G:pn}
\ee
where the integration contour goes around the origin $\lambda=0$,
without encircling any singularity of $\G(\lambda)$.
However, this formula requires a precise knowledge of $\G(\lambda)$,
which we will not be able to arrive at here.
But the behavior of the generating function as a function of $\lambda$
and of the parameter $\Delta x$ will enable us to conjecture
properties of the particle number distribution.

It is clear that when $\lambda\rightarrow 1$ for fixed $\Delta x$,
the generating function $\G(\lambda)$ gets
close to the unweighted
sum over $n$ of $\pn$, and thus by unitarity, tends to~1.
On the other hand, since $\G(\lambda)=p_1(\Delta x) \lambda+\cdots$,
it obviously
tends to $0$ when $\lambda\rightarrow 0$.
From the values of $\lambda$ at which the transition between
0 and 1 occurs, one may
deduce the values of the particle numbers which are the most
probable. Indeed, in the limit in which
the typical particle numbers
are large (which is necessarily relevant when $\Delta x$ is very large),
\be
\G(\lambda)\underset{\Delta x\gg 1}
\simeq \int_0^\infty dn\, e^{-n(1-\lambda)}\,\pn,
\label{eq:pn:G}
\ee
which shows that, under reasonable assumptions
for the probabilities $\pn$,
only values of the particle numbers $n$ less than
$1/(1-\lambda)$ may contribute
significantly to the integral.\footnote{
  The same relations as~(\ref{eq:G:pn}),(\ref{eq:pn:G})
  exist of course between $\pnO$ and $\GO(\lambda)$.
  }

\subsection{\label{sec:numerics}Numerical evaluation
  of the generating functions}

Besides the analytical analysis,
we shall also solve numerically the FKPP equation obeyed by the
functions $Q$ and $H_\phi$: Once these functions are tabulated, the
generating functions $\G(\lambda)$ can be calculated.

We use an implementation first
proposed by Ebert and Van Saarloos (EvS) \cite{EBERT20001}:
The FKPP equation is discretized in space and time (with
respective steps $\delta x$ and $\delta t$)
in a semi-implicit form, in such a way that the scheme leads to a
close approximation of the continuous FKPP evolution, and is numerically
stable even for very large times.

Concretely, let us call $u(x,t)$ a generic function which obeys
the FKPP equation
\be
\partial_t u=\partial_x^2 u+u-u^2,
\ee
and discretize it in $x$ and $t$.
We label the $N_x$ spatial
sites by the integer $j\in[1,N_x]$. For $j\in[2,N_x-1]$,
we replace the
derivatives by finite differences as follows:
\begin{multline}
\frac{u_j(t+\delta t)-u_j(t)}{\delta t}
=\frac12\frac{u_{j+1}(t)-2 u_j(t)+u_{j-1}(t)}{\delta x^2}\\
+\frac12\frac{u_{j+1}(t+\delta t)-2 u_j(t+\delta t)
  +u_{j-1}(t+\delta t)}{\delta x^2}\\
+\frac12\left[u_j(t+\delta t)+u_j(t)\right]
-u_j(t+\delta t)u_j(t).
\label{eq:discretized_u_1}
\end{multline}
Reshuffling the terms of this equation, we see that
the evolution of $u_j$ from times $t$ to
$t+\delta t$ is the solution to the linear set of equations
\begin{multline}
  -\frac12 u_{j+1}(t+\delta t)
+u_j(t+\delta t)
\left\{1+{\delta x^2}\left[\frac{1}{\delta t}-\frac12
              +u_j(t)\right]
  \right\}
-\frac12 u_{j-1}(t+\delta t)\\
=\frac12 u_{j+1}(t)
-u_j(t)\left[1-{\delta x^2}\left(\frac12+
  \frac{1}{\delta t}\right)\right]
+\frac12 u_{j-1}(t).
\end{multline}
We need a rule for the leftmost and rightmost
$x$-lattice sites. We just extend the previous set of equations
to the sites $j=1$ and $j=N_x$ by assuming
the following: $u_0=1$ and $u_{N_x+1}=0$.
Solving this equation then amounts to inverting a tridiagonal matrix
at each timestep, of size the number $N_x$ of lattice sites,
for which there exist very efficient numerical methods~\cite{press1992}.

In addition, in order to minimize the effects of the finite size
of our spatial lattice, we shift back the front every ${\cal O}(1)$
time step, in such a way that its position on the lattice
be approximately unchanged. In practice, at every integer time $t$,
we replace $u_j(t)$ by $u_{j+k_t}(t)$ for all $j\in [1,N_x-k_t]$,
and set $u_j(t)$ to 0 for $j\in [N_x-k_t+1,N_x]$, where $k_t$ is the
integer part of the measured difference in the position of the front
between times $t-1$ and $t$.

The formulas we shall find below
for the branching Brownian motion will still
be valid, up to a set of constants appearing in the
analytical expressions which will
need to be slightly modified, since the effective
diffusion constant and branching rate of the discretized branching
Brownian motion are no longer strictly unity.
The main parameters are the exponential rate $\gamma_0$
at which the traveling wave
goes to zero at large $x$,
its asymptotic velocity $v_0$ and a diffusion constant~$D$,
which are respectively~$1$,~$2$ and~$1$ in the case of the
branching Brownian motion introduced above.
The main changes are in the function $F_t(z)$ in Eq.~(\ref{eq:FtBBM}),
and in $m_t$ in Eq.~(\ref{eq:mtBBM})
which become
\be
F_t(z)=1-C\times z\,e^{-\gamma_0 z}e^{-z^2/4Dt}
\quad\text{and}\quad
m_t=v_0 t-\frac{3}{2\gamma_0}\ln t.
\ee
The values of $\gamma_0$, $v_0$ and $D$ are found
by looking for solutions of the form $u_j(t)=e^{-\gamma[j \delta x-v(\gamma) t]}$ to
the equation obtained by linearization of
Eq.~(\ref{eq:discretized_u_1}) when $u$ is small. $\gamma_0$ minimizes $v(\gamma)$, i.e.
solves $v'(\gamma_0)=0$, and $v_0= v(\gamma_0)$,
$D=\frac12 \gamma_0 v''(\gamma_0)$.

In practice, we set the number of lattice
sites to $N_x=10^4$, the lattice spacing in $x$ to $\delta x=0.05$, and
the lattice spacing in the $t$-variable to $\delta t=0.01$.
With these latter two choices,
\be
\begin{split}
v_0&\simeq 2+\frac23 \delta t^2+\frac{1}{12}\delta x^2+\cdots=2.000275\cdots\\
\gamma_0&\simeq 1-\frac23 \delta t^2-\frac18\delta x^2+\cdots=0.9996208\cdots\\
D&\simeq 1+3 \delta t^2+\frac12\delta x^2+\cdots=1.00155\cdots
\end{split}
\ee
So the differences with the constants characteristic of the BBM are small.
For completeness, let us mention that $F(z)$ in Eq.~(\ref{eq:asymptotics_F}) is also
modified for negative $z$: It becomes $F(z)\simeq C'\times e^{r z}$, with
\be
r=\sqrt{2}-1+\left(\frac43-\sqrt{2}\right)\delta t^2
+\frac13\left(\frac{11}{8}-\sqrt{2}\right)\delta x^2+\cdots\simeq
\left(\sqrt{2}-1\right)-0.0000407\cdots
\ee

\begin{figure}[ht]
  \begin{center}
    \includegraphics[width=.9\textwidth]{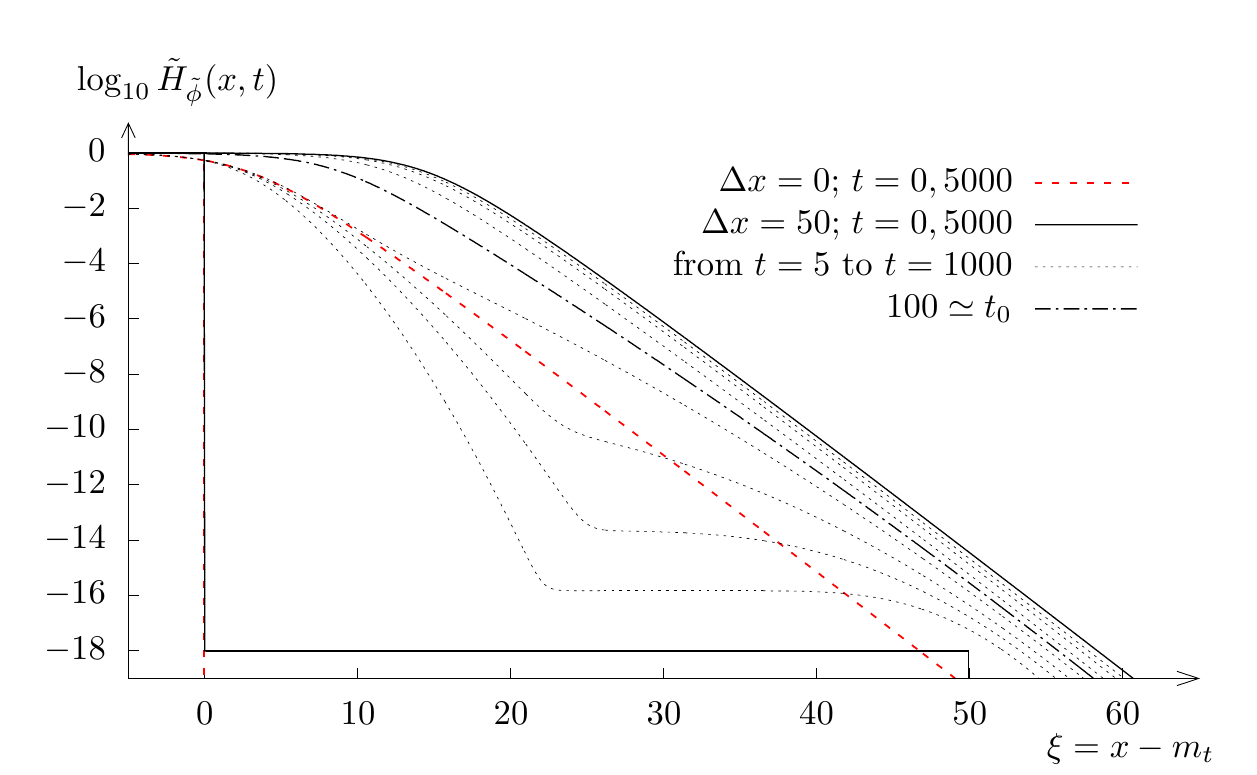}
  \end{center}
  \caption{\small
    Numerical evaluation of $\tilde H_{\tilde\phi}(x,t)$ as a function of $\xi=x-m_t$, that is,
    in the frame of the traveling wave $\tilde H_{\tilde\phi_B}$, at different
    times (logarithmic scale on the vertical axis).
    The parameters that characterize the initial condition
    are set to $\Delta x=50$ and $1-\lambda=10^{-18}$.
    {\it Full line:} Initial condition $\log_{10}\tilde\phi(x)$
    and $\log_{10}\tilde H_{\tilde\phi}(x,t=5000)$, namely at
    a time at which the traveling wave has reached its asymptotic shape.
    The {\it dotted lines} represent
    $\log_{10}\tilde H_{\tilde\phi}(x,t)$
    for $t=5,10,20,50,200,500,1000$ to show the
    evolution between the initial condition and the universal traveling wave.
    The time $t=100$, which is close to the special
    time $\tau$ computed from the formula~(\ref{eq:t0_def}) 
    (see the discussion in Sec.~\ref{sec:smallshift} below)
    is singled out with the help of a {\it dash-dotted line}.
    {\it Dashed lines:}  $\log_{10}\tilde\phi_B(x)$
    and $\log_{10}\tilde H_{\tilde\phi_B}(x,t=5000)$, for comparison.
    \label{fig:plotshape}
  }
\end{figure}

The evolution of $\tilde H_{\tilde\phi}$ from an initial condition characterized
by $\Delta x=50$ and $1-\lambda=10^{-18}$ to the large time
$t=5000$ is displayed in Fig.~\ref{fig:plotshape}. At the final time, the comparison to
$\tilde H_{\tilde\phi_B}$ is shown and a shift $g(\lambda,\Delta x)$ of order
$\Delta x-f(\lambda)\simeq 12$ is clearly visible.
(Observe that the rightmost full curve, representing $\tilde H_{\tilde\phi}$
at a large time, is related
to the dashed curve, representing $\tilde H_{\tilde\phi_B}$ at the same time,
by a mere translation along the $\xi$-axis.)

Once the functions $\tilde H_{\tilde\phi}$ and $\tilde H_{\tilde\phi_B}$ are known for different
values of $\Delta x$, $\psi_\lambda$ and $Q$ follow
from Eqs.~(\ref{eq:H=psi}) and~(\ref{eq:Q=HphiB});
$\GO(\lambda)$ and $\G(\lambda)$ can then be computed from the exact
formulas~(\ref{eq:def_G0}) and~(\ref{eq:def_G}) respectively.
However, since we are interested in the $t\rightarrow +\infty$ and $|x-m_t|\gg 1$ limits,
it is easier to measure numerically
the front positions $m_t$ for some large time, deduce
the shift function $g(\lambda,\Delta x)$ and then
derive \ $\GO(\lambda)$ and $\G(\lambda)$ using Eqs.~(\ref{eq:G0fromg})
and~(\ref{eq:Gfromg}) respectively.
In order to better approach the infinite-time limit, we can extrapolate the finite-$t$
measurements of $g(\lambda,\Delta x)$
by assuming that the leading corrections\footnote{%
  This is because the terms that vanish with $t$ in the large-$t$ expansion
  of $m_t$ form a series in powers of $t^{-1/2}$. The first term, of
  order $t^{-1/2}$ has a universal coefficient (independent of the initial
  conditions)~\cite{EBERT20001};
  Therefore, the first time-dependent correction
  in $g(\lambda,\Delta x)$, which is the difference of the position
  of two FKPP fronts starting from distinct initial conditions,
  is expected to be of order $1/t$.
  } vanish as $\sim 1/t$~\cite{Brunet2011}.
Extrapolating in this way from different times enables us to check the stability
of the procedure.


\section{\label{sec:shift}Asymptotics of the shift and properties of
the generating functions}

\subsection{General strategy}

It will prove useful to change function $H\rightarrow 1-H$.
Therefore, for a generic function ${\cal F}(z)$ taking
values between~0 and~1, we introduce the following notation:
\be
\tilde {\cal F}(z)\equiv 1-{\cal F}(z).
\ee
Then the function $\tilde\phi$, which is the initial condition for the
time evolution of $\tilde H_{\tilde\phi}=1-H_\phi$ when $\phi$ is defined as
in Eq.~(\ref{eq:decomp_phi}), is just the sum of the
functions $\tilde\phi_A$ and $\tilde\phi_B$.

Let us decompose $\tilde H_{\tilde\phi}$ as the sum of two functions
corresponding
to the evolution either of $\tilde\phi_A$ or of $\tilde\phi_B$.
We may perform this
decomposition in two ways. First, we write
\be
\tilde H_{\tilde\phi}=\tilde H_{\tilde\phi_B}+\Delta \tilde H_{\tilde\phi_A}.
\ee
If we require the first term in the r.h.s. to obey the FKPP equation with
$\tilde\phi_B$ as an initial condition, then, since $\tilde H_{\tilde\phi}$ also
obeys the FKPP equation, the functions $\tilde H_{\tilde\phi_B}$ and
$\Delta\tilde H_{\tilde\phi_A}$
solve the following hierarchical system of equations:
\be
\left\{
\begin{split}
\partial_t \tilde H_{\tilde\phi_B}(x,t)&=\partial_x^2 \tilde H_{\tilde\phi_B}(x,t)
+\tilde H_{\tilde\phi_B}(x,t)-\tilde H_{\tilde\phi_B}^2(x,t)\\
\partial_t \Delta \tilde H_{\tilde\phi_A}(x,t)&=
\partial_x^2 \Delta \tilde H_{\tilde\phi_A}(x,t)
+\left[1-2\tilde H_{\tilde\phi_B}(x,t)\right]\Delta \tilde H_{\tilde\phi_A}(x,t)
-\Delta\tilde H_{\tilde\phi_A}^2(x,t).
\end{split}
\right.
\label{eq:decompB}
\ee
The solution of the first equation reads, in the limit of infinite time:
\be
\tilde H_{\tilde\phi_B}(\xi+m_t,t)\underset{t\rightarrow +\infty}\simeq
\tilde F(\xi)\equiv 1-F(\xi).
\label{eq:HphiBsol}
\ee
$\Delta \tilde H_{\tilde\phi_A}$ obeys an equation with a new term with
respect to the FKPP equation, proportional to $\tilde H_{\tilde\phi_B}$.
This term vanishes when $\tilde H_{\tilde\phi_B}\ll 1$, while
it damps the linear growth
term when $\tilde H_{\tilde\phi_B}\sim 1$.
This way of decomposing the equation for $\tilde H_{\tilde\phi}$ will prove
useful when $\tilde H_{\tilde\phi_B}$ eventually dominates $\tilde H_{\tilde\phi}$,
and $\Delta\tilde H_{\tilde\phi_A}$ can be considered a small perturbation.

But we can also write
\be
\tilde H_{\tilde\phi}=\tilde H_{\tilde\phi_A}+\Delta \tilde H_{\tilde\phi_B},
\ee
where
\be
\left\{
\begin{split}
\partial_t \tilde H_{\tilde\phi_A}(x,t)&=\partial_x^2 \tilde H_{\tilde\phi_A}(x,t)
+\tilde H_{\tilde\phi_A}(x,t)-\tilde H_{\tilde\phi_A}^2(x,t)\\
\partial_t \Delta \tilde H_{\tilde\phi_B}(x,t)&=
\partial_x^2 \Delta \tilde H_{\tilde\phi_B}(x,t)
+\left[1-2\tilde H_{\tilde\phi_A}(x,t)\right]\Delta \tilde H_{\tilde\phi_B}(x,t)
-\Delta\tilde H_{\tilde\phi_B}^2(x,t).
\end{split}
\right.
\label{eq:decompA}
\ee
Here, $\tilde H_{\tilde\phi_A}$ obeys the FKPP equation, but with the initial
condition $\tilde\phi_A$, which is a rectangle of width $\Delta x$
and height $1-\lambda$. Its large-time solution is also a traveling
wave $\tilde F$, but shifted by a function of $\lambda$ and $\Delta x$.
Since $\tilde H_{\tilde\phi_A}$ starts out very small, $\Delta\tilde H_{\tilde\phi_B}$
dominates in the beginning of the evolution, and thus this decomposition may seem
irrelevant. However, we will see that there are cases in which
$\tilde H_{\tilde\phi_A}$ eventually dominates
while $\Delta\tilde H_{\tilde\phi_B}$ can be considered a small perturbation,
and those will turn out to be the most interesting cases.

All functions appearing in these equations take values between~0 and~1.
Both systems of equations~(\ref{eq:decompB}),(\ref{eq:decompA})
linearize and decouple
in regions in which $\tilde H_{\tilde\phi}\ll 1$. In these regions,
the functions $\tilde H_{\tilde\Phi}$, $\Delta\tilde H_{\tilde\Phi}$ (with
$\tilde\Phi=\tilde\phi_A$ or $\tilde\phi_B$) approximately
obey an equation of the form
\be
\partial_t \tilde h(x,t)=\partial_x^2 \tilde h(x,t)+\tilde h(x,t),
\label{eq:linearized}
\ee
where $\tilde h$ stands for a solution to Eq.~(\ref{eq:linearized})
constrained by the appropriate
initial condition.

This equation can
easily be solved. However, even in the regions in which the linear
approximation is valid, the nonlinearities may act as
effective absorptive boundary conditions, deforming the 
solution of the initial-value problem.
It is known that by putting an absorptive boundary on the linearized equation,
one can get an approximate solution to the FKPP equation, which converges to
the exact solution at large time~(see e.g. Ref.~\cite{EBERT20001}). Our approach will essentially
consist in replacing the nonlinearities appearing
in Eqs.~(\ref{eq:decompB}),(\ref{eq:decompA}) by suitable boundary conditions.

The first step is to solve Eq.~(\ref{eq:linearized}) with the initial condition
$\tilde\phi_A$ and an absorptive (moving) boundary to effectively represent the
nonlinearities. Once this is done, we will be able
to compute $\tilde H_{\tilde\phi}$
in limiting cases in which the $\Delta\tilde H_{\tilde\Phi}$ functions
in the decompositions bring very small contributions
to  $\tilde H_{\tilde\phi}$ in the limit of large times.

\subsection{Linear evolution of a flat rectangle}

\subsubsection{Free boundaries}

Let us solve the linearized equation~(\ref{eq:linearized}).
We start with the localized initial condition $\delta(x-x_0)$ at some
positive time $t_0$, without imposing boundary conditions at this
stage.
The exact solution at time $t>t_0$ reads
\be
\tilde h_{x_0}(x,t-t_0)=\frac{1}{\sqrt{4\pi (t-t_0)}}e^{t-t_0}
\exp\left[-\frac{(x-x_0)^2}{4(t-t_0)}\right].
\label{eq:htilde0}
\ee
Let us go to a frame comoving with
the traveling wave solution to the FKPP equation for $Q$,
by changing coordinates
$x\rightarrow \xi=x-m_t$, $x_0\rightarrow \xi_0=x_0-m_{t_0}$,
and let us keep only the leading terms for $\xi\ll{t}$.
In the two cases which will be relevant to what follows,
\be
  \tilde h_{\xi_0+m_{t_0}}(\xi+m_t,t-t_0)=
  \begin{cases}\frac{t}{\sqrt{4\pi}}e^{-(\xi-x_0)}
    \exp\left[-\frac{(\xi-x_0)^2}{4t}\right] &
  \text{if $t_0=0$},\\
  \frac{1}{\sqrt{4\pi(t-t_0)}}
  \left(\frac{t}{t_0}\right)^{3/2}e^{-(\xi-\xi_0)}
  \exp\left[-\frac{(\xi-\xi_0)^2}{4(t-t_0)}\right] &
  \text{if $t_0\gg 1$},
  \end{cases}
\label{eq:htilde0comoving}
\ee
where it is understood that $m_{t=0}=0$; The large-time asymptotical
formula~(\ref{eq:mtBBM}) has also been used for $m_t$, as well as for $m_{t_0}$
in the case $t_0\gg 1$.

We now turn to the case in which the initial condition
$\tilde\phi_A(x)$ is taken at $t_0=0$.
This case will prove relevant since
the solution to the linearized equation~(\ref{eq:linearized})
with this initial condition
is an approximation to the solution to the full nonlinear equations
for $\tilde H_{\tilde\phi_A}$ valid when $\tilde H_{\tilde\phi_A}\ll 1$,
and for $\Delta\tilde H_{\tilde\phi_A}$
when $\Delta\tilde H_{\tilde\phi_A}\ll 1$ and $\tilde H_{\tilde\phi_B}\ll 1$
simultaneously. We get $\tilde h_{\tilde\phi_A}$
by superposing the solutions with initial conditions localized at $x_0$,
see Eq.~(\ref{eq:htilde0}),
for all $x_0\in[0,\Delta x]$, and by weighting them uniformly
by $1-\lambda$:
\be
\tilde h_{\tilde\phi_A}(x,t)=\int_0^{\Delta x}
dx_0\,(1-\lambda)\,\tilde h_{x_0}(x,t)
=\frac12 (1-\lambda){e^{t}}\left[\erfc \frac{x-\Delta x}{\sqrt{4t}}
  -\erfc \frac{x}{\sqrt{4t}}\right],
\label{eq:Htilde1}
\ee
where $z\mapsto\erfc(z)$ is the complementary error
function, defined with the following
normalization~\cite{abramowitz1965handbook}:
\be
\erfc(z)\equiv\frac{2}{\sqrt{\pi}}\int_z^{+\infty}dw\,e^{-w^2}.
\ee
We will eventually be
interested in the shift $g(\lambda,\Delta x)$,
which we get focusing on the region of $x$
where $\tilde h_{\tilde\phi_A}\sim{\cal O}(1)$, namely
for $x\sim 2t$.
In this region, obviously, $x\gg\sqrt{t}$, and $\Delta x$ is finite,
such that both error functions
may be expanded using
\be
\erfc (z)\underset{z\rightarrow+\infty}{\sim} \frac{e^{-z^2}}{z\sqrt{\pi}}.
\ee
Since furthermore $\Delta x\gg 1$, only the $\Delta x$-dependent
term in Eq.~(\ref{eq:Htilde1}) is relevant. Hence
\be
\tilde h_{\tilde\phi_A}(x,t)\simeq \frac{1-\lambda}{\sqrt{\pi}}
\frac{\sqrt{t}}{x-\Delta x}
e^{t}\exp\left[-\frac{(x-\Delta x)^2}{4t}\right].
\ee
Going to the comoving frame defined above and keeping again
only the leading terms for $\xi\ll{t}$, we find
\be
\tilde h_{\tilde\phi_A}(\xi+m_t,t)\simeq\frac{1}{\sqrt{4\pi}}\,(1-\lambda)\,
t\, e^{-(\xi-\Delta x)}\exp\left[-\frac{(\xi-\Delta x)^2}{4t}\right].
\label{eq:Htilde2}
\ee
We note that the following identity holds in these approximations:
\be
\tilde h_{\tilde\phi_A}(\xi+m_t,t)=(1-\lambda)\,\tilde h_{x_0=\Delta x}(\xi+m_t,t).
\label{eq:ident}
\ee
In words, the rectangular initial condition leads asymptotically
to the same traveling wave as an initial condition localized at the
rightmost edge of the rectangle, namely at
$x=\Delta x$.

Let us stress that
$\tilde h_{\tilde\phi_A}$ becomes of order 1 when
\be
   {\cal E}\equiv -\ln\frac{1}{1-\lambda}+\ln t-(\xi-\Delta x)
   -\frac{(\xi-\Delta x)^2}{4t}
\ee
goes to zero.
The equation ${\cal E}=0$ has a solution if $t>t_-$, where $t_-$ obeys
\be
t_-+\ln t_-=\ln\frac{1}{1-\lambda}.
\ee
By iteration, keeping only the large terms when $\lambda\rightarrow 1$, we
can get a closed expression for $t_-$:
\be
t_-\simeq f(\lambda),\quad\text{with}\quad
f(\lambda)\equiv\ln\frac{1}{1-\lambda}-
\ln\ln\frac{1}{1-\lambda}.
\label{eq:shiftBD}
\ee
For $t>t_-$, the rightmost point for which unitarity
is reached is at the position
\be
\xi=\Delta x-2t\left[1-\sqrt{1-\frac{1}{t}
    \left(\ln\frac{1}{1-\lambda}-\ln t\right)}\right].
\ee


\subsubsection{Generic absorptive boundary}

We now add an absorptive boundary condition to the linear equation, which
will later be appropriately chosen to represent the relevant nonlinearities.
Let us denote the position of the boundary in the comoving frame by $X_t$.
Mathematically, the absorptive boundary condition reads
\be
\tilde h(x=X_t+m_t,t)=0.
\ee
The solution to this new problem is obtained through the method
of images.\footnote{%
  The method of images would be straightforward if the boundary were enforced
  at a fixed $x$. Here, the position of the boundary has a $t$-dependence, so
  a little more care is needed: See, for example, Ref.~\cite{Mueller:2002zm} for
  the first application of this method in the context of high-energy physics, to
  derive the solution of an equation for
  the hadronic scattering amplitudes at very high energies.
}

We start with the localized
initial condition $\delta(x-x_0)$ at time $t_0$.
The method of images
amounts to modifying  Eq.~(\ref{eq:htilde0comoving})
by replacing the Gaussian factor therein as follows:
\be
\exp\left[-\frac{(\xi-\xi_0)^2}{4(t-t_0)}\right]
\longrightarrow
\exp\left[-\frac{(\xi-\xi_0)^2}{4(t-t_0)}\right]
-\exp\left[-\frac{(\xi+\xi_0-2X_t)^2}{4(t-t_0)}\right].
\ee
The latter can be rewritten as
\be
\exp\left[-\frac{(\xi-X_t)^2}{4(t-t_0)}\right]
\left\{
\exp\left[\frac{(\xi_0-X_t)(\xi-X_t)}{2(t-t_0)}\right]
-\exp\left[-\frac{(\xi_0-X_t)(\xi-X_t)}{2(t-t_0)}\right]
  \right\}
  \exp\left[
    -\frac{(\xi_0-X_t)^2}{4(t-t_0)}
    \right].
\ee
For $t$ such that none of the two Gaussian factors give a large suppression,
i.e. for $|\xi_0-X_t|$ and $|\xi-X_t|$ small compared to $t-t_0$,
which, we anticipate, can both be realized by taking $t$
large enough and by choosing
appropriately $\xi$, one may linearize the
difference of exponentials. This results in the following expression:
\be
\tilde h_{\xi_0+m_{t_0},X_t}(\xi+m_t,t-t_0)\simeq
\begin{cases}
  \frac{1}{\sqrt{4\pi}}\,
(x_0-X_t)(\xi-X_t)\,e^{-\xi+x_0}
  \exp\left[-\frac{(\xi-X_t)^2}{4t}-\frac{(x_0-X_t)^2}{4t}\right]
  & \text{if $t_0=0$},\\
 \frac{1}{\sqrt{4\pi}}\left[\frac{t}{t_0(t-t_0)}\right]^{3/2}\,
(\xi_0-X_t)(\xi-X_t)\,e^{-\xi+\xi_0}\\
\hfill\times\exp\left[-\frac{(\xi-X_t)^2}{4(t-t_0)}
   -\frac{(\xi_0-X_t)^2}{4(t-t_0)}\right] & \text{if $t_0\gg 1$},
 \end{cases}
\label{eq:boundary0}
\ee
where we kept the Gaussian factors as cutoffs.

We turn to the case of the rectangular initial condition $\tilde\phi_A$
at $t_0=0$. Actually, we may get
the answer without any further calculation, by taking advantage of
the identification~(\ref{eq:ident}).
For reasons that will become clear below,
we introduce $\delta_t\equiv \Delta x-X_t$ in order to
parametrize the position of the boundary.
Simple replacements in Eq.~(\ref{eq:boundary0}) lead to
\be
\tilde h_{\tilde\phi_A,\delta_t}(\xi+m_t,t)
\simeq\frac{1}{\sqrt{4\pi}}(1-\lambda)\,
\delta_t\,(\xi-\Delta x+\delta_t)\,
e^{-\xi+\Delta x}
\exp\left[-\frac{(\xi-\Delta x+\delta_t)^2}{4t}-\frac{\delta_t^2}{4t}\right].
\label{eq:Htilde3}
\ee

In the present context,
we will basically replace the nonlinearities in Eqs.~(\ref{eq:decompB})
and~(\ref{eq:decompA}) by appropriate moving boundaries.
The latter equations boil then down to the linear equation~(\ref{eq:linearized})
that governs
the evolution of $\tilde h$, supplemented by boundary conditions.


\subsection{\label{sec:limiting}Limiting cases}

\subsubsection{\label{sec:smallshift}Small shift}

If $\lambda\rightarrow 1$
while $\Delta x$ is kept fixed,
one does not expect that $\tilde H_{\tilde\phi}$ be very different from
$\tilde H_{\tilde\phi_B}$.
In this case, it seems appropriate to
address the problem starting from
the system of equations~(\ref{eq:decompB}).
$\Delta\tilde H_{\tilde\phi_A}$
in that decomposition can be regarded
as a perturbation. Consequently, the shift $g(\lambda,\Delta x)$
will be small for such values
of the parameters. (We will soon determine the parametric region
in which this is true.)

The function
$\tilde H_{\tilde\phi_B}(x,t)$ obeys the ordinary FKPP equation, and thus its
solution is a traveling wave front at position $x=m_t$, i.e. at $\xi=0$
by definition of the comoving frame; See Eq.~(\ref{eq:HphiBsol}).
In the initial stages of the evolution and for $\xi>0$,
the equation for $\Delta\tilde H_{\tilde\phi_A}$ can be thought of as a linear
equation with an absorptive boundary at the position~$\xi=0$ of the front $B$.
Hence $\Delta\tilde H_{\tilde\phi_A}\propto\tilde h_{\tilde\phi_A,\delta_t=\Delta x}$,
and thus, from  Eq.~(\ref{eq:Htilde3}),
\be
\Delta\tilde H_{\tilde\phi_A}(\xi+m_t,t)\simeq C_0\times(1-\lambda)\Delta x \,
e^{\Delta x-\Delta x^2/4t}
\times\xi\, e^{-\xi-\xi^2/4t}.
\label{eq:DeltaHsmall}
\ee
We do not have control over the overall factor $C_0$,
which we however expect to be of order 1.

We note that $\Delta\tilde H_{\tilde\phi_A}$ remains of order 1 or
less for any $\xi>0$
throughout the time evolution whenever the parameters
$\Delta x$ and $\lambda$ satisfy
\be
(1-\lambda)\Delta x \,e^{\Delta x}<1,\quad\text{namely when}\quad
\Delta x<f(\lambda),
\label{eq:cdts}
\ee
where $f(\lambda)$ is
the time $t_-$ introduced in Eq.~(\ref{eq:shiftBD}).

Let us assume that the condition~(\ref{eq:cdts}) is strongly satisfied,
i.e. $f(\lambda)-\Delta x\gg 1$,
so that $\Delta\tilde H_{\tilde\phi_A}\ll 1$ throughout and for all $\xi>0$.
Then the second nonlinear term in
its evolution equation~(\ref{eq:decompB}) can always be neglected.
Therefore, in this limit, the infinite-time solution of the system of equations
reads
\be
\begin{split}
  \tilde H_{\tilde\phi_B}(\xi+m_t,t)&\underset{\text{Eq. (\ref{eq:HphiBsol}),
      $t\rightarrow+\infty$}}{\simeq}C\times\xi e^{-\xi},\\
  \Delta \tilde H_{\tilde\phi_A}(\xi+m_t,t)&
  \underset{\text{Eq. (\ref{eq:DeltaHsmall}),
      $t\rightarrow+\infty$}}{\simeq}
C_0\times(1-\lambda)\Delta x \,e^{\Delta x}\times\xi\, e^{-\xi}.
\end{split}
\ee
The sum of these two functions amounts to $\tilde H_{\tilde\phi}$,
which as we know becomes, in the infinite-time limit, the same traveling wave
as $\tilde H_{\tilde\phi_B}$ but shifted by $g(\lambda,\Delta x)$:
\be
\begin{split}
\tilde H_{\tilde\phi_B}+\Delta \tilde H_{\tilde\phi_A}&=\tilde H_{\tilde\phi}\\
\implies C\times\xi e^{-\xi}\left[
  1+\frac{C_0}{C}\times(1-\lambda)\Delta x \,e^{\Delta x}
  \right]&\simeq C\times[\xi-g(\lambda,\Delta x)] e^{-\xi+g(\lambda,\Delta x)}.
\end{split}
\ee
This equality should hold for $\xi\gg 1$. Since
$g(\lambda,\Delta x)$ is small compared to 1, taking $\xi$ much larger than~1,
the $\xi$-dependence
cancels left and right. This equation turns into a closed expression
for $g(\lambda,\Delta x)$, up to the unknown constant $c\equiv C_0/C$:
\be
g(\lambda,\Delta x)\simeq\ln\left[
  1+c(1-\lambda)\Delta x\, e^{\Delta x}
  \right]\simeq c(1-\lambda)\Delta x\, e^{\Delta x}.
\label{eq:g_small}
\ee
This expression is valid
when the parameters satisfy the condition~(\ref{eq:cdts}) strongly.
The generating functions $\GO(\lambda)$ and
$\G(\lambda)$ are then very close to~1.

According to Eq.~(\ref{eq:G0fromg}) with $g(\lambda,\Delta x)$
replaced by the expression in Eq.~(\ref{eq:g_small}),
sticking to the $\Delta x\gg 1$ limit,
$\GO(\lambda)$ reads
\be
\GO(\lambda)
\simeq 1-(1-\lambda)\,c\,\Delta x\, e^{\Delta x}.
\label{eq:G0smallshift}
\ee
According to Eqs.~(\ref{eq:Gfromg}) and~(\ref{eq:g_small}),
$\G(\lambda)$ reads
\be
  \G(\lambda)\simeq 
\begin{cases}
1-(1-\lambda) c'\, e^{\Delta x}
& \text{for $x>m_t$}\\
1-(1-\lambda) c\sqrt{2}\,\Delta x\, e^{\Delta x}
& \text{for $x<m_t$}.
\end{cases}
\label{eq:Gsmallshift}
\ee
$c'$ is another constant, a priori distinct from $c$, that we cannot determine
due to
our lack of control of the terms of order $(1-\lambda)e^{\Delta x}$ in Eq.~(\ref{eq:g_small}).

The explicit
expressions~(\ref{eq:G0smallshift}) and~(\ref{eq:Gsmallshift})
will enable us to
derive {\it mean} particle numbers in the interval $[x-\Delta x,x]$, see
Sec.~\ref{sec:mean} below.

\begin{figure}[ht]
  \begin{center}
    \includegraphics[width=0.9\textwidth]{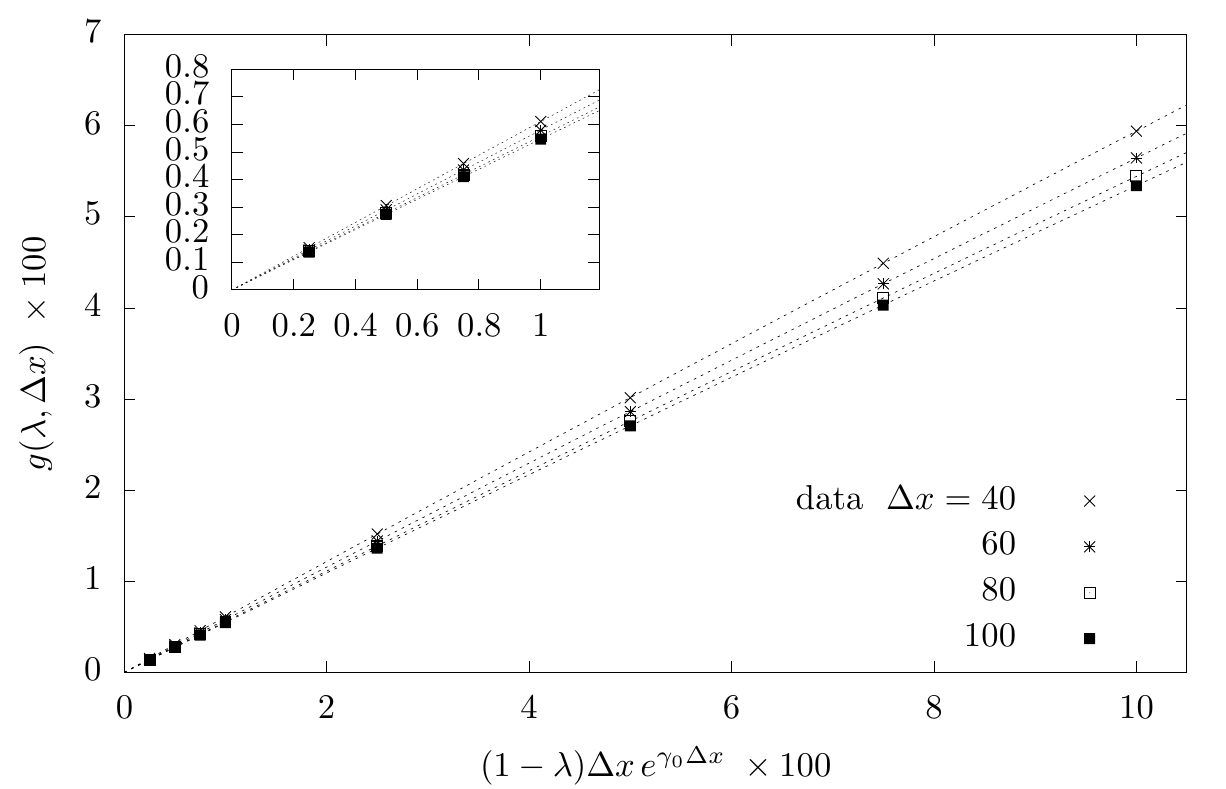}
    \end{center}
\caption{\label{fig:smallshift}{\small
  Asymptotic shift $g(\lambda,\Delta x)$ as a function of the variable
  $(1-\lambda)\Delta x\,e^{\gamma_0\Delta x}$, for different values of $\Delta x$.
  Note that this variable, as well as the shift, are both scaled by a factor 100, for
  the sake of a better lisibility.
  The {\it inset} is a zoom, by a factor~10,
  on the small values of the variable. The {\it dotted lines}
  represent quadratic fits to the numerical data for fixed $\Delta x$,
  intended to guide the eye.
  The slopes at the origin
  for the two largest values of $\Delta x$ differ by about 2\%.
  }
}
\end{figure}

\paragraph{Comparison to a numerical solution of the FKPP equation.}
We can easily check these formulas for the generating functions numerically.
It is actually enough to check the form~(\ref{eq:g_small}) for
the shift\footnote{We actually include the factor $\gamma_0$ in the exponent
in Eq.~(\ref{eq:g_small}) in order to take into account the difference
between the FKPP equation
and its discretized version we are solving, see the caption
of Fig.~\ref{fig:smallshift}.} $g(\lambda,\Delta x)$:
The expressions for the generating functions indeed follow from the one for the shift
through the exact relations~(\ref{eq:Gsmallshift}),(\ref{eq:G0smallshift})
in the infinite-time limit.

Hence, for different choices of $\lambda$ and $\Delta x$ such that
$f(\lambda)-\Delta x\gg 1$, we solve the FKPP equation starting from the
initial condition $\phi$, and we compare,
at some large time $t$,
the position of the obtained front with that of the front that results from
the evolution of the initial condition $\phi_B$.
(Actually, in practice, we measure $g$ at time $t=9000$, and again at time $t=12000$,
and use the two obtained values to extrapolate to $t=+\infty$ under the assumption
that the difference to the asymptotical value decreases as $1/t$. We checked
that we get indeed very close to the asymptotics by comparing to extrapolations
from smaller times.)

Our numerical results are displayed in Fig.~\ref{fig:smallshift}: They show
a good agreement with the formula~(\ref{eq:g_small}), which improves as $\Delta x$
grows larger and $\lambda$ gets closer to~$1$.

\subsubsection{Large shift}
  
If $\Delta x>f(\lambda)$ instead, then the linearized evolution drives
$\Delta\tilde H_{\tilde\phi_A}$
to values larger
than~1. We observe that this happens at a time on the order of
\be
\tau\equiv\frac{\Delta x^2}{4[\Delta x-f(\lambda)]}.
\label{eq:t0_def}
\ee
$\tau$ turns out to be the main timescale in this problem: It is actually
the time at which
the evolution of $A$ reaches unitarity at the position of the front $B$.
When this happens,
we cannot regard  $\Delta\tilde H_{\tilde\phi_A}$ as a perturbation to
$\tilde H_{\tilde\phi_B}$. Instead, the shift $g(\lambda,\Delta x)$
of the position of the
asymptotic traveling wave
will be substantial.
In this situation, it may be more appropriate to address the problem
from the point of view of
the system of equations~(\ref{eq:decompA}).
Indeed, we expect that the evolution of
the front $A$ brings a significant contribution to $\tilde H_{\tilde\phi}$.
Therefore, we shall study the limit in which, at large times, the front $A$ dominates
$\tilde H_{\tilde\phi}$, while the front $B$ is a small perturbation.

The first equation of the system~(\ref{eq:decompA}) is the FKPP
equation for $\tilde H_{\tilde\phi_A}$. For times less than 
$t_-$ defined in Eq.~(\ref{eq:shiftBD}),
the evolution is essentially linear. For larger times, we replace
the nonlinearity by a moving absorptive
boundary at $\Delta x-\delta_t$, as in Eq.~(\ref{eq:Htilde3}):
$\tilde H_{\tilde\phi_A}\propto\tilde h_{\tilde\phi_A,\delta_t}$.
Now, we set $\delta_t$ in such a way that
the maximum of $\tilde H_{\tilde\phi_A}$ be~1.
A straightforward calculation leads to the following expression for $\delta_t$:
\be
\delta_t=2t\left(1-\sqrt{1-\frac{1}{t}\ln \frac{1}{1-\lambda}}\right)
-\frac{1}{\sqrt{1-\frac{1}{t}{\ln \frac{1}{1-\lambda}}}}
  \ln\left[2t\left(
  1-\sqrt{1-\frac{1}{t}\ln \frac{1}{1-\lambda}}
  \right)\right].
\ee
Expanding for large $t$ and $\lambda\rightarrow 1$,
\be
\delta_t=\left(\ln\frac{1}{1-\lambda}-\ln\ln\frac{1}{1-\lambda}\right)
+\frac{1}{4t}\ln^2 \frac{1}{1-\lambda}
+{\cal O}\left(\frac{1}{t}\ln\frac{1}{1-\lambda}
\ln\ln\frac{1}{1-\lambda}\right).
\ee
In particular, $\delta_{t\rightarrow\infty}=f(\lambda)$,
where $f(\lambda)$ was defined in Eq.~(\ref{eq:shiftBD}).
The asymptotic front is then described by
\be
\begin{split}
  &\tilde H_{\tilde \phi_A}(x=\xi+m_t,t)\underset{t\rightarrow+\infty}{\simeq}
     C\times[\xi-\Delta x+f(\lambda)]
       \,e^{-\xi+\Delta x-f(\lambda)}\\
&  \text{with}\quad
       f(\lambda)=\ln\frac{1}{1-\lambda}
       -\ln\ln\frac{1}{1-\lambda}.
\end{split}
\label{eq:asymptoticHA}
\ee
These expressions are valid in the limit $t\rightarrow+\infty$, $x-m_t\gg 1$,
$|\ln(1-\lambda)|\gg 1$, $\xi-\Delta x+f(\lambda)\gg 1$.
We have recovered a result first obtained by
Brunet and Derrida~\cite{Brunet_2009} for the ``delay''
of a traveling wave when the initial condition for the FKPP equation
is the small step $(1-\lambda)\Theta(\Delta x-x)$,
with respect to a traveling wave evolved from the full step function $\Theta(-x)$
connecting 0 and 1.

Let us now turn to the discussion of $\Delta\tilde H_{\tilde\phi_B}$.
So long as $t< \tau$, where $\tau$ is given by Eq.~(\ref{eq:t0_def}),
its evolution
is driven by the plain FKPP equation: Indeed, the term involving
$\tilde H_{\tilde\phi_A}$ in the second equation
of the system~(\ref{eq:decompA}) can be neglected. This means that
\be
\Delta\tilde H_{\tilde\phi_B}(\xi+m_t,t)\underset{t<\tau}{\simeq}
C\times\xi\,e^{-\xi-\xi^2/4t}.
\label{eq:earlyDeltaH}
\ee
But for $t>\tau$, the term involving
$\tilde H_{\tilde\phi_A}$ in Eq.~(\ref{eq:decompA}) is no longer negligible:
Front $A$ starts to ``cut off'' front $B$.
The evolution of $B$ ahead of this effective cutoff
may then be approximated by the linearized
equation supplemented by an absorptive boundary
at the position of front~$A$, namely at $\xi=\Delta x-f(\lambda)$.
But the size of front $B$ at the time $\tau$ at which the evolution
of $A$ starts to cut it off is of order $\sqrt{\tau}$, see Eq.~(\ref{eq:earlyDeltaH}),
that is $\Delta\tilde H_{\tilde\phi_B}$ is negligible
in the region $\xi>\sqrt{\tau}$.
Consequently, if
\be
\sqrt{\tau}\ll\Delta x-f(\lambda),
\quad\text{namely}\quad
\Delta x-f(\lambda)\gg\Delta x^{2/3},
\ee
then the evolution of $\tilde H_{\tilde\phi}$ coincides
with a good approximation with the evolution of
front~$A$: for $t>\tau$,
$\tilde H_{\tilde\phi}\simeq\tilde H_{\tilde\phi_A}$.
In this case, the asymptotic shift is simply the position
of front $A$ in the comoving frame, at infinite time, namely
\be
g(\lambda,\Delta x)\underset{\Delta x-f(\lambda)\gg(\Delta x)^{2/3}}{\simeq}
\Delta x-f(\lambda).
\ee

If instead $1\ll\Delta x-f(\lambda)\ll\Delta x^{2/3}$, then, 
at time $\tau$, front $B$ is cut off by front $A$ in a region
in which $\tilde H_{\tilde\phi_B}\sim 1$.
Thus for $\xi\gg\Delta x-f(\lambda)$, $\tilde H_{\tilde\phi}$
is just the sum of two
FKPP fronts
\be
\tilde H_{\tilde\phi}(\xi+m_t,t)\underset{t\rightarrow +\infty}{\simeq}
 C\times\xi
 \,e^{-\xi+\Delta x-f(\lambda)}
 +C\times\xi\,e^{-\xi}.
 \label{eq:BperturbsA}
 \ee
 In the present limit in which $\Delta x-f(\lambda)$ is large,
 it is clear that the second
 term (front $B$) is small compared to the first one (front $A$):
 Front $A$ eventually
 dominates, while front $B$ can be considered a perturbation.
 The shift is calculated by solving the equation
 $\tilde H_{\tilde\phi}[\xi-g(\lambda,\Delta x)+m_t,t]=\tilde H_{\tilde\phi_B}(\xi+m_t,t)$ in the limit
of infinite time and large $\xi$.
We find
\be
g(\lambda,\Delta x)=\Delta x-f(\lambda)+ e^{-\Delta x+f(\lambda)}.
\ee

Introducing a function $\alpha(\lambda,\Delta)$, we may summarize
the properties of the shift in the region $\Delta x>f(\lambda)$
by the following formula:
\be
\begin{split}
g(\lambda,\Delta x)=&\Delta x-f(\lambda)+\alpha(\lambda,\Delta x)\,
e^{-\Delta x+f(\lambda)}\\
&\text{with}\quad
\begin{cases}
  \alpha(\lambda,\Delta x)\rightarrow 1&
  \text{when $\Delta x-f(\lambda)\ll\Delta x^{2/3}$}\\
  \alpha(\lambda,\Delta x)\rightarrow 0&
  \text{when $\Delta x-f(\lambda)\gg\Delta x^{2/3}$}
\end{cases}
\end{split}
\label{eq:g_general}
\ee
Using Eqs.~(\ref{eq:G0fromg}) and~(\ref{eq:Gfromg})
to relate the shift to the
generating functions, we find that $\GO(\lambda)$ and
$\G(\lambda)$ read
\be
\begin{cases}
\GO(\lambda)=\left[
  \alpha(\lambda,\Delta x)-\frac{\partial\alpha(\lambda,\Delta x)}
        {\partial\Delta x}
        \right]
\exp\left[-\Delta x+f(\lambda)\right]\\
  \G(\lambda)=
\left[
  \alpha(\lambda,\Delta x)-\frac{\partial\alpha(\lambda,\Delta x)}
        {\partial\Delta x}
        \right]
\exp\left[\alpha(\lambda,\Delta x)e^{-\Delta x+f(\lambda)}\right].
\end{cases}
\label{eq:GandG0intermediate}
\ee
It is clear from these formulas that
$\G(\lambda)$ is of
order~1 for any $(\lambda,\Delta x)$ satisfying
$1\ll\Delta x-f(\lambda)\lesssim\Delta x^{2/3}$
(and actually also when the first inequality is released),
while it tends to zero for $\Delta x-f(\lambda)\gg\Delta x^{2/3}$.
This is in sharp contrast with $\GO(\lambda)$ for which the
parametric region where it is of order~1 is reduced to
$\Delta x-f(\lambda)\lesssim 1$, because of the exponential
factor.


\subsection{\label{sec:intermediate}
  Intermediate region: Scaling law for the generating function
$\G(\lambda)$}

Let us now try and pin down more properties of the function $\alpha$
introduced in Eq.~(\ref{eq:g_general}).

We have identified $\tau$ defined in Eq.~(\ref{eq:t0_def})
as a relevant time scale: It is the time at which
front $A$ starts to cut off front $B$.
Since the position
of front~$A$ eventually reads $\Delta x-f(\lambda)$, see Eq.~(\ref{eq:asymptoticHA}),
the square of the latter quantity is another relevant time scale:
It is the time needed for a FKPP front to extend over that
distance. We define
\be
t_\text{diff}\equiv \frac14\left[\Delta x-f(\lambda)\right]^2.
\label{eq:tdiff_def}
\ee
It is then natural to expect the generating function $\G(\lambda)$
to scale with the ratio $\sigma^2\equiv t_\text{diff}/\tau$, namely
\be
\text{$\G(\lambda)$ is a function of}
\ \ \sigma^2(\lambda,\Delta x)\equiv\frac{[\Delta x-f(\lambda)]^{3}}{\Delta x^2}
\ \ \text{only}.
\label{eq:scaling}
\ee

A full calculation of $G_{\Delta x}(\lambda)$ in the relevant parametric region
would be out of reach.
However, as will be explained in the next subsection~\ref{sec:scalingtoy},
such a scaling emerges from an analytical calculation
in the context of a toy model for the system of equations~(\ref{eq:decompA}),
and can be searched for numerically, as will be shown
in subsection~\ref{sec:scalingnum}.

\subsubsection{\label{sec:scalingtoy}Scaling in a toy model}

In the parametric regime of interest in this section,
the shift function is essentially determined by the $A$ front, while the
$B$ front, at large time, brings a small perturbation to the latter.
What makes however this case complicated is that in the beginning of the
evolution, until a time $t_0\sim \tau$, the evolution is dominated by front $B$;
Namely, $\tilde H_{\tilde\phi}(x,t)\underset{t<t_0}\simeq\Delta\tilde H_{\tilde\phi_B}(x,t)$.
This is because $\tilde H_{\tilde\phi_A}$ starts out very small,
and remains so in the region $\xi>0$
until $t_0$.

As discussed above, for $t$ much larger than $t_0$
and in the region $\xi-[\Delta x-f(\lambda)]\gg 1$,
the time evolution of $\Delta \tilde H_{\tilde\phi_B}$ can be approximated by a
linear equation with a boundary
at the position where the nonlinearity in the equation for
$\tilde H_{\tilde\phi_A}$
cuts off the linear evolution,
namely at $\xi=\Delta\equiv\Delta x-f(\lambda)$ in the moving frame,
which is, to a good accuracy, the
position of the asymptotic front.
In order to be able to conduct a complete calculation,
we shall take the model assumption that the transition occurs instantaneously at
time $t_0$.
Therefore, we first write
\be
\Delta\tilde H_{\tilde\phi_B}(\xi+m_{t_0},t_0)= \tilde H_{\tilde\phi_B}(\xi+m_{t_0},t_0)=
1-F_{t_0}(\xi)\underset{\text{Eq.~(\ref{eq:FtBBM})}}{\simeq}
C\times\xi\, e^{-\xi-\xi^2/4t_0}.
\label{eq:DeltaHt1}
\ee
At time $t$ such that $t-t_0\gg{\cal O}(\sqrt{t_0})$,
since $\Delta\tilde H_{\tilde\phi_B}(\xi,t_0)$ results from the evolution
of the linearized FKPP equation with a boundary that moves at
the velocity of a FKPP front, it must have the following shape in the
region $\xi>\Delta$:
\be
\Delta\tilde H_{\tilde\phi_B}(\xi+m_t,t)\simeq
c_\Delta\times\xi_\Delta e^{-\xi_\Delta-\xi_\Delta^2/4t}
  \quad\text{with}\quad
  \xi_\Delta\equiv\xi-\Delta.
  \label{eq:DeltaH_generic}
\ee
The only new unknown in this formula is the constant $c_\Delta$,
that we shall express as a function of $C$, $\Delta$ and~$t_0$.

Since our simplified model consists technically
in solving a linear equation with boundaries,
we may use a completeness relation at $t_0$:
Each point at coordinate $\xi_0>\Delta$ at time $t_0$ evolves
independently into a front through the linearized equation
supplemented with an absorptive boundary at position $\Delta$
in the comoving frame, and
the resulting front is the superposition of all these partial fronts
weighted by $\Delta\tilde H_{\tilde\phi_B}(\xi_0+m_{t_0},t_0)$.
In equations,
\be
\Delta\tilde H_{\tilde\phi_B}(\xi+m_t,t)=
\int_\Delta^\infty
d\xi_0\,\Delta\tilde H_{\tilde\phi_B}(\xi_0+m_{t_0},t_0)
\times\tilde h_{\xi_0+m_{t_0},\Delta}(\xi+m_t,t-t_0)
\ee
Using Eq.~(\ref{eq:Htilde3}) and Eq.~(\ref{eq:DeltaHt1}),
and going to the $t\rightarrow+\infty$ limit while
keeping $t_0$ and $\Delta$ fixed,
\be
\Delta\tilde H_{\tilde\phi_B}(\xi+m_t,t)
\underset{t\rightarrow+\infty}{=}
\frac{C}{\sqrt{4\pi}}\times e^{-\Delta}
\,\xi_\Delta e^{-\xi_\Delta}
\left[\frac{1}{t_0^{3/2}}
\int_\Delta^\infty d\xi_0\,
(\xi_0-\Delta)
\xi_0 e^{-\xi_0^2/4t_0}
\right].
\ee
The $\xi_0$-integral can be performed exactly:
One may notice e.g. that the factor
$\xi_0\,e^{-\xi_0^2/4t_0}$ may also be written as
$-2t_0\times \frac{d}{d\xi_0}e^{-\xi_0^2/4t_0}$,
and integrate by parts.
We arrive at the elegant expression
\be
\Delta\tilde H_{\tilde\phi_B}(\xi+m_t,t)={C}\times e^{-\Delta}
\erfc\frac{\Delta}{2\sqrt{t_0}}
\,\xi_\Delta\, e^{-\xi_\Delta}.
\label{eq:DeltaH_toy}
\ee
The constant $c_\Delta$ is readily identified by comparison
of Eq.~(\ref{eq:DeltaH_toy}) with Eq.~(\ref{eq:DeltaH_generic})
in the $t\rightarrow+\infty$ limit.

In order to match this simplified calculation with our
initial problem, we replace
$\Delta$ by $\Delta x-f(\lambda)$, and set
\be
t_0=\kappa^2 \tau,
\ee
where $\kappa$ is a number of order~1: This arbitrary constant
takes into account the fact that in the initial problem, the effective
absorptive boundary
does not exhibit a discontinuity at
time $\tau$. Instead, the transition occurs smoothly in a time window
of size of order $\tau$ from $t=\tau$.
Then,
\be
\frac{c_\Delta}{C} e^{\Delta}=
\erfc\frac{[\Delta x-f(\lambda)]^{3/2}}{\kappa\Delta x}.
\label{eq:cDelta_scaling}
\ee

As in the case of a small shift discussed in Sec.~\ref{sec:smallshift},
the function $g(\lambda,\Delta x)$ is obtained by writing
$\tilde H_{\tilde\phi}$ as the sum of $\tilde H_{\tilde\phi_A}$ and
$\Delta\tilde H_{\tilde\phi_B}$,
and by computing the position of the resulting traveling wave
in the reference frame comoving with the front
$\tilde H_{\tilde\phi_B}$.
Keeping only the leading terms, we arrive at the following expression:
\be
g(\lambda,\Delta x)
\simeq\Delta x-f(\lambda)+
e^{-[\Delta x-f(\lambda)]}
  \erfc
  \frac{[\Delta x-f(\lambda)]^{3/2}}{\kappa\Delta x}.
\ee
Again, we can deduce an expression for the generating function $\G(\lambda)$ by using
Eq.~(\ref{eq:Gfromg}). The latter takes a very simple form when
$1\ll[\Delta x-f(\lambda)]^{3/2}\lesssim\Delta x$:
\be
\G(\lambda)\simeq\erfc\frac{[\Delta x-f(\lambda)]^{3/2}}{\kappa\Delta x}.
\label{eq:G_scaling}
\ee
We do not expect this one-free-parameter
simple functional form to be exact, since it was derived
in a simplified model for the actual system of equations from which
$\G(\lambda)$ was deduced. However,
it may be a valuable expression to start with, and in particular,
to compare with numerical solutions to the FKPP equation.

\subsubsection{\label{sec:scalingnum}Numerical check of the scaling}

We can compute numerically $\G(\lambda)$ for different values of the
parameters $\Delta x$ and $\lambda$, and check the scaling~(\ref{eq:G_scaling}).
We employ the method based on the measurement of the shift function $g(\lambda,\Delta x)$
explained in Sec.~\ref{sec:numerics}.

Although the numerical evaluation of $g(\lambda,\Delta x)$
is in principle straightforward, it
is not so easy in practice to get $\G(\lambda)$ to a fair accuracy. Indeed, we need to
evolve the FKPP equation to large times to make sure that the traveling wave
has approached its
asymptotic shape. We set the maximum evolution time to~$8000$ in all
our calculations.
A further difficulty is that for large $g(\lambda,\Delta x)$, the formula~(\ref{eq:Gfromg}) 
becomes the product of a large term, the exponential, by a small prefactor, which
is numerically awkward. This puts limits on the range of the parameters
we are able to explore with our current implementation: The largest value of
$\Delta x$ we shall consider
is 80, and we shall stick to the region $\sigma\lesssim {\cal O}(1)$.

\begin{figure}
  \begin{center}
    \includegraphics[width=.9\textwidth]{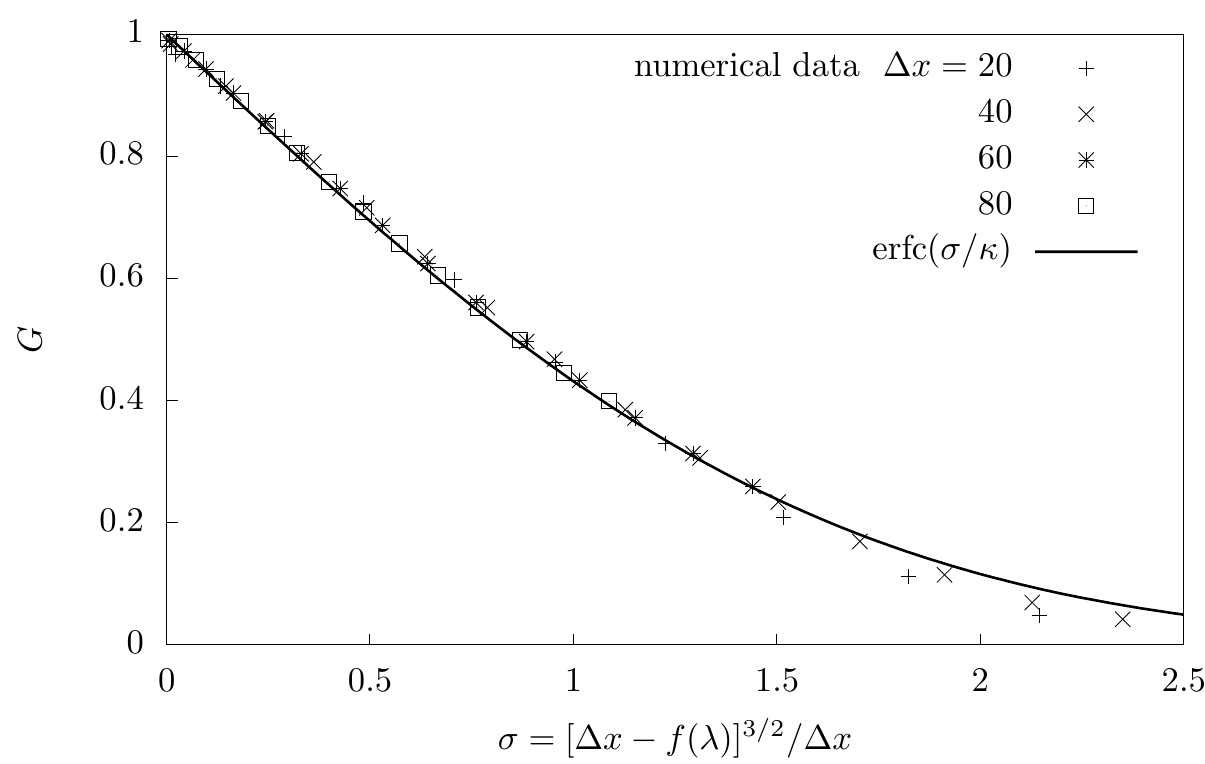}
  \end{center}
  \caption{\small \label{fig:plotscalingG}
    Generating function $\G(\lambda)$ extracted from the
    numerical solution of the FKPP equation as a function
    of the scaling variable
    ${[\Delta x-f(\lambda)]^{3/2}}/{\Delta x}$.
    The numerical data is compared
    to Eq.~(\ref{eq:G_scaling}) with $\kappa$ set to 1.8.
    }
\end{figure}

The result of the numerical calculation is displayed in Fig.~\ref{fig:plotscalingG}:
The generating function is plotted against the scaling variable $\sigma$, where
for the latter we used the definition in Eq.~(\ref{eq:scaling}) with an additional
overall factor
$1/\gamma_0$ in $f(\lambda)$ meant to take into account the effect of the discretization,
see the discussion in Sec.~\ref{sec:numerics}.
We estimate that the numerical accuracy is on the order of the size of the data points.

The numerical data
exhibit the scaling~(\ref{eq:scaling})
quite spectacularly, especially given that the values of $\Delta x$
for which we are able to produce data barely allow
to satisfy the constraints $1\ll\Delta x-f(\lambda)\lesssim \Delta x^{2/3}$
which we would expect necessary for the scaling to be verified.
Interestingly enough,
the formula~(\ref{eq:G_scaling}) for $\G(\lambda)$ established in the context
of the toy model reproduces fairly well the numerical data,
with the parameter $\kappa$ arbitrarily set to the value $1.8$,
which, as expected, is indeed of order unity.


\section{\label{sec:distribution}Particle numbers in the tip}

In this section, we apply the results we got for the shift $g(\lambda,\Delta x)$
and for the generating function $\G(\lambda)$
to the characterization of the particle number distribution
within a distance $\Delta x$ from the lead particle.

\subsection{\label{sec:mean}Mean particle numbers from small shifts}

The expected number of particles
\be
{\bar n}(\Delta x)\equiv n^{(1)}(\Delta x)=\sum_n n\,\pn
\ee
in an interval of size $\Delta x$ from the
lead particle can be obtained from the coefficient of the
term of order $\lambda-1$ in the $\lambda\rightarrow 1$ expansion
of the generating function $\GO(\lambda)$ (if the position of the
lead particle is not constrained, namely if one
considers typical events) or $\G(\lambda)$ (if the position of the
lead particle is fixed).
The expressions~(\ref{eq:Gsmallshift}) and~(\ref{eq:G0smallshift})
are accurate enough to enable us to read off these expectation values:
\begin{itemize}
\item The mean number of particles in typical realizations
  of the BBM (namely in which the lead particle is not constrained to be
  at a given position)
  can be calculated
  from $\GO(\lambda)$.
  Identifying Eq.~(\ref{eq:G0smallshift}) to the two first terms in
  Eq.~(\ref{eq:G_expansion}), we find, in the limit $\Delta x\gg 1$
  in which our analytical expressions are expected to be exact:
  \be
     {\bar n}(\Delta x)\simeq c\,\Delta x\, e^{\Delta x}.
     \label{eq:meantypical}
  \ee
  We have recovered a result already
  obtained by Brunet and Derrida in Ref.~\cite{Brunet2011}.

\item The mean number of particles when the position of the
  lead particle is fixed
  is deduced from $\G(\lambda)$ in Eq.~(\ref{eq:Gsmallshift}).
  Two cases must be distinguished:
  \begin{itemize}
  \item The position of the lead particle, $x$,
    is chosen far to the left of its expectation value,
    i.e. $m_t-x\gg 1$. Then, comparing Eqs.~(\ref{eq:Gsmallshift})
    and~(\ref{eq:G_expansion})
            we get
  \be
  {\bar n}(\Delta x)\simeq c\sqrt{2}\,\Delta x\, e^{\Delta x}.
  \ee

\item Finally, in the case of main interest in this paper, when
  the lead particle is unusually ahead its expected position,
  $x-m_t\gg 1$, we get
  \be
  {\bar n}(\Delta x)\simeq c'\,e^{\Delta x}.
  \ee
  \end{itemize}

\end{itemize}

We note that ${\bar n}(\Delta x)$ is smaller
in the last case than in the case when one averages over the position
of the lead particle, but only by a factor $\Delta x$.
The leading factor, $e^{\Delta x}$, is the same.
Since ${\bar n}(\Delta x)$ is computed by weighting $\pn$ by $n$, the configurations
which contribute most are the ones which have a large number of particles
in the tip. Requiring $n$ to be large is tantamount to asking for
a robust front extending at least from $x-\Delta x$ to $x$, which
in the phenomenological model \cite{Mueller:2014gpa}, can only be generated by a
front fluctuation if $\Delta x$ is large.
Therefore, we expect the number of particles
in a typical tip fluctuation to be less than ${\bar n}(\Delta x)\propto e^{\Delta x}$.


\subsection{Particle distributions in rare tip fluctuations: Tentative estimate}

Let us now investigate the consequences of the scaling~(\ref{eq:scaling})
of the generating function that we have found.

\subsubsection{Particle numbers}

Let us define $\lambda_0$ and $\lambda_\zeta$ as solutions of the equations
\be
\Delta x-f(\lambda_0)=0
\quad\text{and}\quad
\Delta x-f(\lambda_\zeta)=\zeta\,\Delta x^{2/3},
\ee
where $\zeta$ is a constant of order unity.
The explicit solutions read
\be
\frac{1}{1-\lambda_0}\simeq\Delta x\,e^{\Delta x}
\quad\text{and}\quad
\frac{1}{1-\lambda_\zeta}\simeq\Delta x\,e^{\Delta x-\zeta\Delta x^{2/3}}.
\ee
$\lambda_0$ is the point where the scaling variable $\sigma$ is $0$,
while $\lambda_\zeta$ is the point at which $\sigma=\zeta^{3/2}$.
According to the discussion in Secs.~\ref{sec:limiting} and~\ref{sec:intermediate}
(see also Fig.~\ref{fig:plotscalingG}),
the transition region in which $\G(\lambda)$ drops significantly from~1
coincides with the parametric region $0<\sigma\lesssim\zeta^{3/2}$.
For fixed~$\Delta x$, this corresponds to $\lambda$
describing the interval $[\lambda_\zeta,\lambda_0]$.
According to Eq.~(\ref{eq:pn:G}), the probability $\pn$ is
significant for $1/(1-\lambda_\zeta)<n<1/(1-\lambda_0)$, namely for
\be
\Delta x\, e^{\Delta x-\zeta\,\Delta x^{2/3}}<n<\Delta x\, e^{\Delta x}.
\label{eq:pn_tip}
\ee

Note that the situation is very different for the number of particles
in an interval $\Delta x$ from the lead particle in {\it typical} realizations.
It is easy to see from the formula in Eq.~(\ref{eq:GandG0intermediate})
that the generating function of the corresponding probabilities,
$\GO(\lambda)$,
has its transition region around $1/(1-\lambda)$ of the order of
$\Delta x\, e^{\Delta x}$: More precisely, this region roughly
extends
over the $\lambda$-interval $[\lambda_0',\lambda_0]$, where
$\lambda_0'$ solves $\Delta x-f(\lambda_0^\prime)=1$.
This implies that the typical number of particles in an interval
of size $\Delta x$ from the lead particle in typical realizations
coincides, in order of magnitude, with the mean number of particles,
see Eq.~(\ref{eq:meantypical}).

Equation~(\ref{eq:pn_tip}) shows that the particle number
may be smaller by a factor on the order of $e^{\zeta\Delta x^{2/3}}$ in rare tip
fluctuations than it would be in typical tips.
This would be enough for the main assumption of
the phenomenological model
for fluctuations in branching random walks to prove correct:
The latter indeed assumes that the particle density in tip fluctuations is less
than $e^{\Delta x}$.
However, the dominant behavior of typical particle numbers
in the limit of large interval sizes
is still given by the exponential $e^{\Delta x}$, which may seem
a bit surprising.

\subsubsection{Mean distances between nearby particles}

Let us recall that
Brunet and Derrida computed rigorously the large-$n$ asymptotics
of the mean distances between particles number $n$ and $n+1$ from
the lead particle when the position of the latter is left
free. The result reads~\cite{Brunet_2009}
\be
\langle d_{n,n+1}\rangle=\frac{1}{n}-\frac{1}{n\ln n}.
\ee
They observed that these distances are smaller than in the case
of a Poisson process of exponential intensity $e^x$, for which
$\langle d_{n,n+1}\rangle=1/n$. However, they turn out to
coincide with what would be found in the case of a Poisson process
on a line of intensity $x\, e^{x}$~\cite{Brunet2011}.

Let us come back to rare tip fluctuations.
If, in these realizations,
the particles were distributed according to a Poisson process
on a line of intensity $x\,e^{x-\zeta\, x^{2/3}}$, then
the mean distance between particle number $n$ and particle number $n+1$
from the lead particle would read
\be
\langle d_{n,n+1}\rangle=\frac{1}{n}+\frac{2\zeta}{3}\frac{1}{n(\ln n)^{1/3}}.
\ee
We see that this distance is {\it larger} than that found both in
the case in which the lead particle is not an unusual fluctuation,
and in the case of the Poisson process with exponential intensity.
However, we know that there should exist strong correlations,
especially in the rare realizations we select,
which would a priori
invalidate a Poisson process assumption.


\section{Conclusions and outlook}

With the aim of understanding tip fluctuations in branching random walks,
our main thrust was the study, in the $t\rightarrow+\infty$ limit,
of the generating function $\G(\lambda)$ of the particle-number
probabilities $\pn$ to observe $n$ particles in an interval of size $\Delta x$
from the lead particle, the position $x$ of the latter being fixed to some large
number, in such a way that the difference between $x$ and its expectation value $m_t$
be large. In this limit, the $x$-dependence of $\G(\lambda)$
vanishes. Our main result can be summarized as follows for the branching Brownian motion:
\be
\text{Defining}\ \sigma^2\equiv\frac{[\Delta x-f(\lambda)]^3}{\Delta x^2},
\quad G_{\Delta x}(\lambda)=\text{func}
\left(\sigma^2\right)\\
\simeq
\begin{cases}
  1&\text{for $\Delta x<f(\lambda)$}\\
    {\cal O}(1)&\text{for $\sigma\sim 1$}\\      
      0&\text{for $\sigma\gg 1$}
\end{cases}
\ee
where $f(\lambda)$ is the Brunet-Derrida ``delay function'' \cite{Brunet_2009}
\be
f(\lambda)=\ln\frac{1}{1-\lambda}-\ln\ln\frac{1}{1-\lambda}\ .
\ee
We gave analytical arguments in support of this scaling,
and checked it numerically, see Fig.~\ref{fig:plotscalingG}.
Proving this scaling and finding an accurate expression
for $\G(\lambda)$ are outstanding challenges.

We have also found that the expectation value of the particle number grows like
$e^{\Delta x}$ with the size of the interval. However, mean particle numbers
are probably dominated by front fluctuations, at variance with typical particle numbers,
which we expect to build up in late stages of the evolution.

The scaling for the generating function $\G(\lambda)$
has enabled us to estimate heuristically the typical values of the number of particles
in a rare tip fluctuation as
\be
\left.n(\Delta x)\right|_\text{typical}\sim \Delta x\,e^{\Delta x-\zeta\Delta x^{2/3}}.
\ee
If this guess is confirmed, the phenomenological model for fluctuations
in branching random walks is indeed
justified, since this number is much
less than $e^{\Delta x}$. However, it lies surprisingly
close to the exponential asymptotically in the limit of large $\Delta x$.

Knowing the scaling form of the generating function is not enough
to allow one to fully calculate
the probability $\pn$ of observing $n$ particles in an
interval of given size from the tip when the position of the latter is fixed,
or the mean distance between two nearby
particles. Such quantities would help us to better assess the particle
distribution in these fluctuations.

It was not possible either to calculate the particle number
probabilities~$\pn$ from our numerical implementation: The latter
provides the generating function $\G(\lambda)$, $\pn$ would be related to the
$n$-th $\lambda$-derivative of $\G(\lambda)$, which cannot
be computed in practice for $n$ much larger than one.
A Monte Carlo implementation of the stochastic
process would be more useful, but the latter is not straightforward
since we are interested in extremely
rare realizations of the branching Brownian motion.
(Note that algorithms for generating specifically
rare events have been designed in
a different subfield of statistical physics,
see e.g. Ref.~\cite{Giardina2011}).
We leave this for further investigations.

We also plan to address the correlations of the positions
of the lead particles at different times, which will give
valuable information on the way tip fluctuations develop over time.

\section*{Acknowledgements}

Our  research  is  supported in part by
the U.S. Department of Energy Grant
\# DE-FG02-92ER40699,
and by the Agence Nationale
de la Recherche under the project \# ANR-16-CE31-0019.
We thank the Department of Physics of the University of Florence (Italy)
for hospitality at the time when this work was initiated,
and Professors Stefano Catani and Dimitri Colferai for their
welcome. SM thanks Dr Alexandre Lazarescu for interesting discussions
on arguments closely related to this work.



{\small

}

\end{document}